\documentclass[usenatbib,usegraphicx]{mn2e}
\usepackage{times}
\usepackage{psfig,latexsym,graphicx,natbib} 
\bibpunct{(}{)}{;}{a}{}{,}
\pssilent

\usepackage[svgnames]{xcolor}   % to colour parts of text

\definecolor{beige}{HTML}{FFDF8C}

\begin{document}
\title[Basic parameters and properties of HR\,7355]{Basic parameters and
  properties of the rapidly rotating magnetic helium strong B star
  HR\,7355\thanks{Based on observations collected at ESO's Very Large
    Telescope under Prog-IDs 081.D-2005 and 383.D-0095.}}
\author[Th. Rivinius et al.]
{Th.~Rivinius$^1$\thanks{triviniu@eso.org},
R.H.D.~Townsend$^2$,
O.~Kochukhov$^3$,
S.~\v{S}tefl$^1$,
D.~Baade$^4$,\newauthor
L.~Barrera$^5$,
Th.~Szeifert$^1$
\\
% \offprints{Th.\ Rivinius\\
%\thanks{triviniu@eso.org}}
%
$^1$ ESO - European Organisation for Astronomical Research in the Southern
Hemisphere, Casilla 19001, Santiago 19, Chile \\
$^2$ Department of Astronomy, University of Wisconsin--Madison, Sterling Hall,
475 N Charter St., Madison 53711, USA \\
$^3$ Department of Physics and Astronomy, Uppsala University Box 516, 751 20 Uppsala, Sweden\\
$^4$ ESO - European Organisation for Astronomical Research in the Southern
Hemisphere, Karl-Schwarzschild-Str.~2, 85748 Garching bei M\"unchen, Germany\\
$^5$ Universidad Metropolitana de Ciencias de la Educac\'ion (UMCE), Jos\'e
Pedro Alessandri 774, Santiago, Chile\\
}
\date{Accepted: $<$date$>$; Received: $<$date$>$}
   \maketitle

\begin{abstract}
{The spectral and magnetic properties and variability of the B2Vnp
  emission-line magnetosphere star HR\,7355 were analyzed. The object rotates
  at almost 90\% of the critical value, meaning it is a magnetic star for
  which oblateness and gravity darkening effects cannot be ignored any
  longer. A detailed modeling of the photospheric parameters indicate that the
  star is significantly cooler than suggested by the B2 spectral type, with
  R$_{\rm eff}=17\,500$\,K atypically cool for a star with a helium enriched
  surface.  The spectroscopic variability of helium and metal lines due to the
  photospheric abundance pattern is far more complex than a largely dipolar,
  oblique magnetic field of about 11 to 12\,kG may suggest. Doppler imaging
  shows that globally the most He enriched areas coincide with the magnetic
  poles and metal enriched areas with the magnetic equator. While most of the
  stellar surface is helium enriched with respect to the solar value, some
  isolated patches are depleted.  The stellar wind in the circumstellar
  environment is governed by the magnetic field, i.e.\ the stellar
  magnetosphere is rigidly corotating with the star. The magnetosphere of
  HR\,7355 is similar to the well known $\sigma$\,Ori\,E: the gas trapped in
  the magnetospheric clouds is fairly dense, and at the limit to being
  optically thick in the hydrogen emission.  Apart from a different magnetic
  obliquity, HR\,7355 and the more recently identified HR\,5907 have virtually
  identical stellar and magnetic parameters.}
\end{abstract}

%
%%%%%%%%%%%%%%%%%%%%%%%%%%%%%%%%%%%%%%%%%%%%%%%%%%%%%%%%%%%%%%%%%%%%%%%%%

%%%%%%%%%%%%%%%%%%%%%%%%%%%%%%%%%%%%%%%%%%%%%%%%%%%%%%%%%%%%%%%%%%%%%%%%%
%%%%%%%%%%%%%%%%%%%%%%%%%%%%%%%%%%%%%%%%%%%%%%%%%%%%%%%%%%%%%%%%%%%%%%%%%
\section{Introduction}\label{sec_intro}

The helium-strong star HR\,7355 \citep{2008A&A...482..255R} has recently been
found to host a magnetic field with several kilogauss longitudinal
strength ($<B_Z>$) \citep{rivi7355,oksala}. The rotational period of
$P=0.52$\,d puts it among the shortest period non-degenerate magnetic stars
known, while its $v\sin i \approx 310$\,km\,s\,$^{-1}$ is the highest for any
known non-degenerate magnetic star.

For a long time, rapid rotation (the term ``rapid rotation'' used here as
being at a fraction of well above 50\% of break-up velocity) and the presence
of a strong magnetic field have been regarded as mutually exclusive for early
type stars, and so the sheer existence of this object poses a number of
puzzling questions. Most important is the stellar age vs.\ the expected spin
down time.  Age estimates for HR\,7355 range from 15 to 25\,Myr
\citep{2010A&A...511L...7M}, much longer than the same authors give for the
spin-down timescale, and hence clearly problematic for fossil field
explanations. However, the gravity of $\log g = 3.95$ found by \citet{oksala}
would be more supportive of a younger star.  In any case, HR\,7355 is a star
for which effects like gravity darkening and oblate deformation cannot be
ignored. This means traditional analysis methods to derive stellar
parameters will, at best, give uncertain results, and at worst misleading
ones.

Intriguing is the presence of photospheric abundance nonuniformities,
supporting the result that rotational mixing even in rapid rotators should be
inhibited at such field strengths \citep{2011IAUS..272...14Z}. Not only are
there abundance variations across the stellar surface in HR\,7355, but the
amplitude of the He{\sc i} equivalent width (EW) variations is much larger
than in the similar, though less rapidly rotating, Bp star $\sigma$\,Ori\,E.

For the following work we will briefly review the rotational ephemeris in
Sect.~\ref{sec_ephem}, then introduce the new observations in
Sect.~\ref{sec_obs}. The first aim of this work is to provide stellar
parameters taking into account the rapid stellar rotation, i.e.\ rotational
deformation and gravity darkening, in Sect.~\ref{sec_param}. The spectroscopic
variations are then described and analyzed in detail in Sect.~\ref{sec_photo},
as are the measured properties of the magnetic field in
Sect.~\ref{sec_magfield} and of the magnetosphere in
Sect.~\ref{sec_magnetosphere}.  The results are discussed in
Sect.~\ref{sec_discussion} and finally the conclusions are drawn in
Sect.~\ref{sec_conclusions}.

%%%%%%%%%%%%%%%%%%%%%%%%%%%%%%%%%%%%%%%%%%%%%%%%%%%%%%%%%%%%%%%%%%%%%%%
\subsection{\label{sec_ephem}Ephemeris}

There are several ephemerides and period determinations, out of which only the
one by \citet{2008A&A...482..255R} is clearly obsolete. In order to converge
on one, we combined the data used by \citet{rivi7355} with the photometric
data published by \citet{2010A&A...511L...7M}, and the photometric data kindly
provided by \citet{oksala}. While the period by \citet{rivi7355} did not sort
well the combined photometric data when phased with this period, the 2008
photometry by \citeauthor{oksala} and the 2008/2009 spectroscopy
(\citeauthor{rivi7355} and this work) do have their minima at acommon
phase. As a result, we adopt here the epoch given by \citet{rivi7355} $T_0
({\rm MJD}) = 54\,940.33 $, which is the mid-date of an occultation of the
star by the magnetospheric material, seen in the Balmer line spectroscopy, and
as such very well defined. However, the better period proved to be the value
published by \citeauthor{oksala}, $P=0.5214404(6)$\,d, who could rely on a
time baseline twice as long. The error of the period is such that for the
purposes of this work it can be fully neglected, and we treat the period as a
fixed value.

%%%%%%%%%%%%%%%%%%%%%%%%%%%%%%%%%%%%%%%%%%%%%%%%%%%%%%%%%%%%%%%%%%%%%%%
%%%%%%%%%%%%%%%%%%%%%%%%%%%%%%%%%%%%%%%%%%%%%%%%%%%%%%%%%%%%%%%%%%%%%%%
\section{Observations}\label{sec_obs}

Apart from archival and literature data, which will be referenced where used,
this work is based on high-resolution echelle spectra obtained in 2009 with
UVES \citep{2000SPIE.4008..534D} at the 8.2m Kueyen telescope on Cerro
Paranal. The instrument was used in the dichroic 2 437/760 setting, which
gives a blue spectrum from about 375 to 498\,nm and a red spectrum from about
570 to 950\,nm, with a small gap at 760\,nm. The slit width was 0.8\arcsec,
giving a resolving power of about $R=50\,000$ over the entire spectrum. The
UVES observations were obtained in three service mode observing runs, in April,
July, and September 2009 (see Table~\ref{tab_obs} for the date ranges).

Exposure times were between 120 seconds in April and 30 seconds in July to
September, the latter ones being repeated four times in a row. Since the
period is short we did not average shorter exposures, but treated them as
taken at distinct phases. A few of the blue spectra taken with $t_{\rm
  exp}=120$\,sec were overexposed in better-than-average seeing conditions,
thus we have more red than blue spectra. In total, we have 104 blue and 112
red spectra. The typical signal to noise ratio ($S/N$) of a single spectrum is
about 275 in the blue region and 290 in the red. However, for unknown reasons,
high exposure levels in UVES, even if still in the fully linear regime of the
detector, tend to cause some order-merging and global normalization problems,
i.e.\ high $S/N$ data obtained with UVES often have more residual echelle
order ``wiggles'' than data with a lower exposure level. This prompts for
local re-normalization of the spectra for analysis of any particular line. In
particular, broad spectral features, like Balmer-line wings, are to be treated
with care. Our analysis of those features is cross-checked on the low
resolution FORS1 long-slit data of \citet{rivi7355}.

%%%%%%%%%%%%%%%%%%%%%%%%%%%%%%%%%%%%%%%%%%%%%%%%%%%%%%%%%%%%%%%%%%%%%%%
%%%%%%%%%%%%%%%%%%%%%%%%%%%%%%%%%%%%%%%%%%%%%%%%%%%%%%%%%%%%%%%%%%%%%%%
%%%%%%%%%%%%%%%%%%%%%%%%%%%%%%%%%%%%%%%%%%%%%%%%%%%%%%%%%%%%%%%%%%%%%%%
\begin{table}
\caption[]{\label{tab_obs}Summary of UVES observations}
\begin{center}
\begin{tabular}{cccc}
Run & \multicolumn{2}{c}{date range} & \# of spectra \\
    & (MJD) & civil (UT)    &blue/red arm\\
\hline
A &54923 -- 54951 &2009-04-02 to 04-30& 29/33\\
B &55026 -- 55062 &2009-07-14 to 08-18& 55/59\\
C &55077 -- 55083 &2009-09-03 to 09-09& 20/20\\
\end{tabular}
\end{center}
\end{table}
%%%%%%%%%%%%%%%%%%%%%%%%%%%%%%%%%%%%%%%%%%%%%%%%%%%%%%%%%%%%%%%%%%%%%%%
%%%%%%%%%%%%%%%%%%%%%%%%%%%%%%%%%%%%%%%%%%%%%%%%%%%%%%%%%%%%%%%%%%%%%%%
%%%%%%%%%%%%%%%%%%%%%%%%%%%%%%%%%%%%%%%%%%%%%%%%%%%%%%%%%%%%%%%%%%%%%%%

%%%%%%%%%%%%%%%%%%%%%%%%%%%%%%%%%%%%%%%%%%%%%%%%%%%%%%%%%%%%%%%%%%%%%%%
%%%%%%%%%%%%%%%%%%%%%%%%%%%%%%%%%%%%%%%%%%%%%%%%%%%%%%%%%%%%%%%%%%%%%%%
%%%%%%%%%%%%%%%%%%%%%%%%%%%%%%%%%%%%%%%%%%%%%%%%%%%%%%%%%%%%%%%%%%%%%%% 
\begin{figure}
\includegraphics[angle=270,width=9cm,clip=]{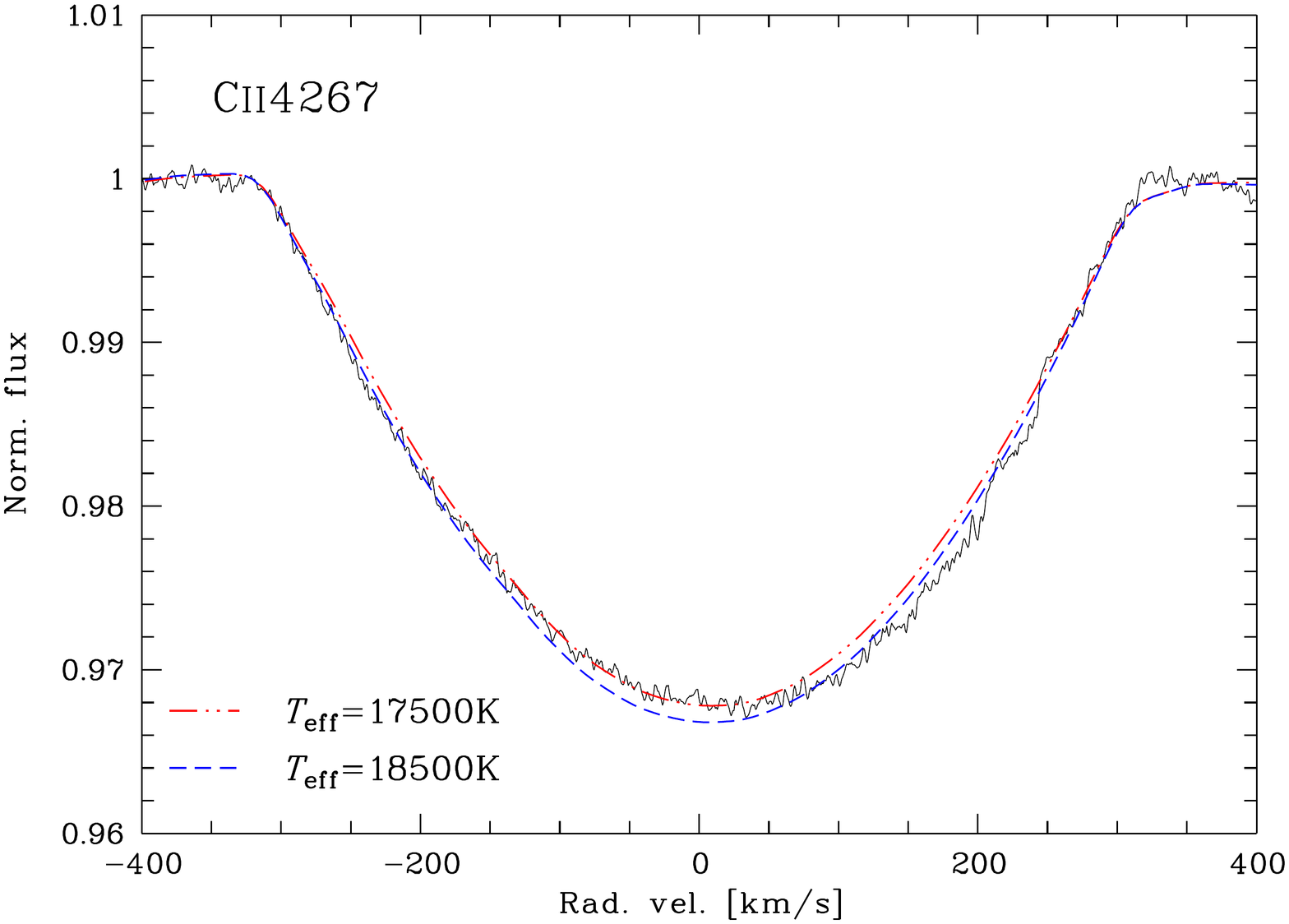}
\caption[]{The averaged UVES line profile for {C}{\sc ii}$\lambda$\,4267 and the best
  fitting model for this particular equivalent width, with $T_{\rm
    eff}=18\,500$\,K, and the final best model, taking into account additional
  conmstraints as discussed in the text, with $T_{\rm eff}=17\,500$\,K, both
  with $v \sin i=310$\,km\,s$^{-1}$ and shifted to $v_{\rm
    sys}=7$\,km\,s$^{-1}$.}
\label{fig_vsini}
\end{figure}
%%%%%%%%%%%%%%%%%%%%%%%%%%%%%%%%%%%%%%%%%%%%%%%%%%%%%%%%%%%%%%%%%%%%%%%
%%%%%%%%%%%%%%%%%%%%%%%%%%%%%%%%%%%%%%%%%%%%%%%%%%%%%%%%%%%%%%%%%%%%%%%
%%%%%%%%%%%%%%%%%%%%%%%%%%%%%%%%%%%%%%%%%%%%%%%%%%%%%%%%%%%%%%%%%%%%%%%

%%%%%%%%%%%%%%%%%%%%%%%%%%%%%%%%%%%%%%%%%%%%%%%%%%%%%%%%%%%%%%%%%%%%%%%
%%%%%%%%%%%%%%%%%%%%%%%%%%%%%%%%%%%%%%%%%%%%%%%%%%%%%%%%%%%%%%%%%%%%%%%
\section{Stellar Parameters}\label{sec_param}
In order to determine the stellar parameters, we made use of spectral
synthesis, both for line strength and profiles, as well as for absolute
fluxes. The code we used is the third, re-programmed version of
\citeauthor{1997MNRAS.284..839T}'s (\citeyear{1997MNRAS.284..839T}) BRUCE and
KYLIE suite, which in the following we will call ``Bruce3''.

Apart from computational issues, and the implementation of perturbations like
pulsation, which is not used here, the main improvement is the treatment of
limb-darkening: the code uses either NLTE atmosphere models by
\citet{2007ApJS..169...83L}, or \citeauthor{1992IAUS..149..225K}'s
(\citeyear{1992IAUS..149..225K}) LTE (ATLAS9) atmospheres, from which the
synthetic spectra were computed by SYNSPEC \citep{2011ascl.soft09022H}, with
wavelength-dependent limb-darkening coefficients, i.e.\ including the effects
of limb-darkening on the spectral lines, ranging from 88 to 750\,nm. Several
characteristics of the line profiles, most notably strengths of C{\sc ii} and
Si{\sc ii} lines in the blue range, were better reproduced with the LTE
atmospheres rather than with the NLTE ones (but still using an NLTE spectral
profile code). The spectrum of HR\,7355 is, therefore, modeled based on an
ATLAS9-SYNSPEC grid.

For line-profile comparison, the synthetic spectra, although in principle
already ``perfectly'' normalized (i.e. according to the computed continuum),
were re-normalized with the same procedure as the observed spectra, i.e.\ with
a series of smoothly joining spline segments, and using the same wavelengths
to sample the continuum.

\subsection{Projected rotational velocity and radius}

In a first step, the averaged profile of {C}{\sc ii}$\lambda$\,4267 was used
to obtain the projected rotational velocity by comparing the profile to the
grid of Bruce3 models. For any reasonable set of parameters, $v \sin i =
310\pm5$\,km\,s$^{-1}$ approximated the width of the observed profile best. We
note that this approach completely neglects macroturbulence or other potential
broadening mechanisms. However, concerning macroturbulence this is a
reasonable approach for a main sequence star, and significant magnetic
broadening is not expected for the measured field. The value is in good
agreement with \citet{oksala} (see Fig.~\ref{fig_vsini}).

The systemic velocity was determined during this measurement as well, to
$v_{\rm sys}=7\pm2$\,km\,s$^{-1}$. We found no indication for a variation of
this value, hence we can exclude a companion close or massive enough to
significantly affect the magnetosphere, unless it would have an orbit in the
plane of the sky, i.e.\ significantly misaligned with the rotation, seen
equatorially (see below).  Assuming rigid rotation of the circumstellar
environment, the derived $v \sin i$ in combination with the rotational period
of $P=0.5214404$\,d the equatorial radius then becomes $R_{\star, \rm eq} \sin
i = 3.19\,{\rm R_{\sun}}$. Assuming that a rapidly rotating star is
constrained by five independent parameters, usually $v_{\rm eq}, i, T_{\rm
  eff}, M_\star, R_{\star, eq}$, this leaves only three of them to be
determined, namely the mass, the effective temperature, and the inclination.

%%%%%%%%%%%%%%%%%%%%%%%%%%%%%%%%%%%%%%%%%%%%%%%%%%%%%%%%%%%%%%%%%%%%%%%
%%%%%%%%%%%%%%%%%%%%%%%%%%%%%%%%%%%%%%%%%%%%%%%%%%%%%%%%%%%%%%%%%%%%%%%
%%%%%%%%%%%%%%%%%%%%%%%%%%%%%%%%%%%%%%%%%%%%%%%%%%%%%%%%%%%%%%%%%%%%%%%
\begin{figure*}
\includegraphics[angle=270,width=18cm,clip=]{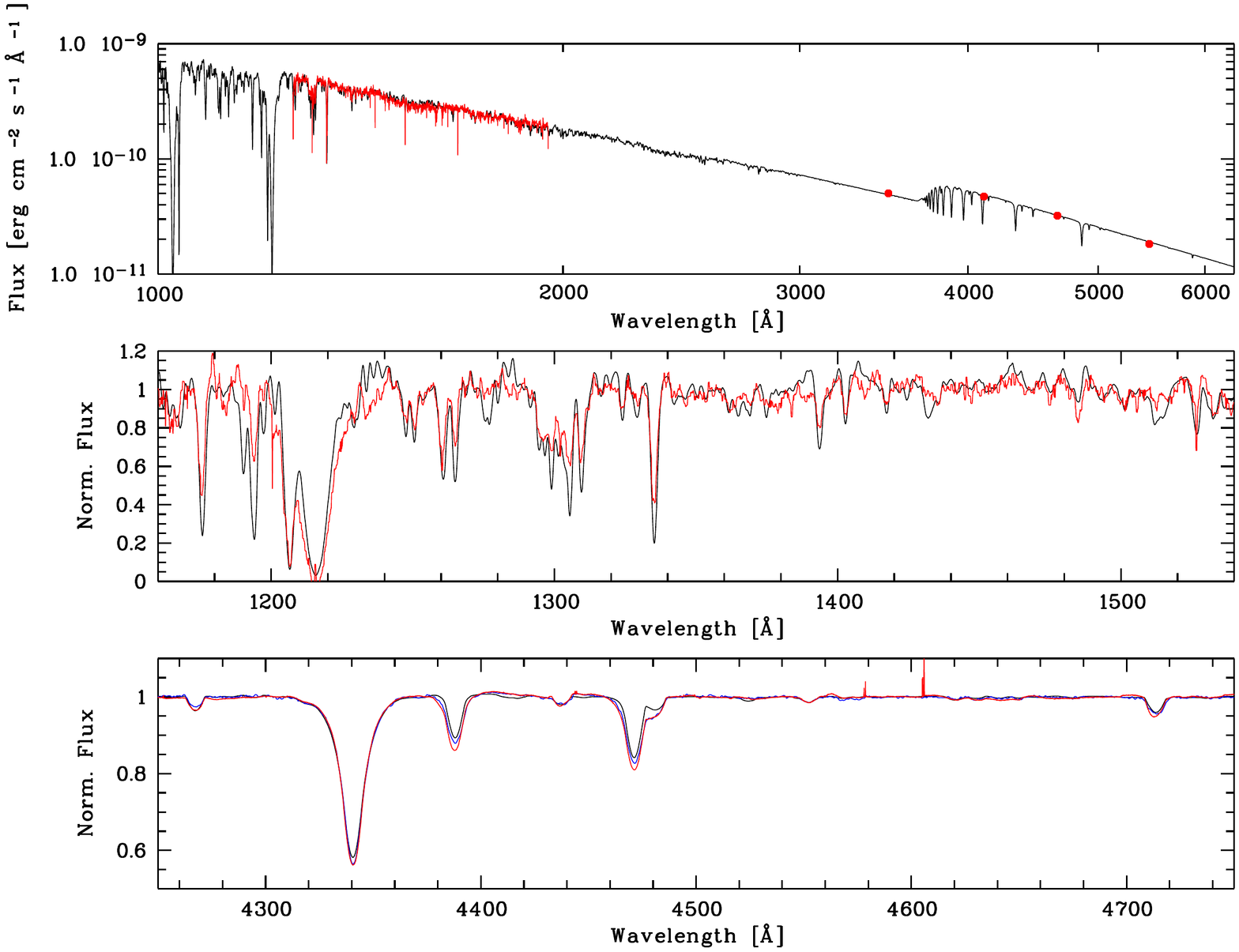}
\caption[]{The Bruce3 model (see Table~\ref{tab_modpar}) vs.\ the observed
  data: Upper panel: SED fit for UV IUE short-wavelength spectrophotometry and
  Str\"omgren visual photometry (large red dots) vs.\ modeled fluxes. Middle
  panel: Line profile fit for IUE spectrum (red) vs.\ modeled UV line profiles
  (black). Lower panel: Line profile fit for averaged UVES (red) and FORS1
  (blue) spectra vs. modeled visual line profiles (black). Note that in this
  step solar abundances are still used, hence the mismatch for the helium
  profiles. This discrepancy is solved later by the Doppler imaging.}
\label{fig_fluxfit}
\end{figure*}
%%%%%%%%%%%%%%%%%%%%%%%%%%%%%%%%%%%%%%%%%%%%%%%%%%%%%%%%%%%%%%%%%%%%%%%
%%%%%%%%%%%%%%%%%%%%%%%%%%%%%%%%%%%%%%%%%%%%%%%%%%%%%%%%%%%%%%%%%%%%%%%
%%%%%%%%%%%%%%%%%%%%%%%%%%%%%%%%%%%%%%%%%%%%%%%%%%%%%%%%%%%%%%%%%%%%%%%

%%%%%%%%%%%%%%%%%%%%%%%%%%%%%%%%%%%%%%%%%%%%%%%%%%%%%%%%%%%%%%%%%%%%%%%
\subsection{Effective temperature}

The interpretation of the effective temperature $T_{\rm eff}$ in a rapidly
rotating star, i.e.\ in a star where gravity darkening and oblateness have
become significant, is not straightforward. So we note that Bruce3 uses
$T_{\rm eff}$ in the sense that it is the uniform temperature, that a
black-body of the same surface area would need to have, to have the same
luminosity as the actual gravity darkened star, integrated over the full solid
angle, while one has to keep in mind that observationally derived $T_{\rm
  eff}$ are usually solid-angle dependent quantities and only assumed to be
valid over the entire sphere. The value for $T_{\rm eq}$ and $T_{\rm pol}$ in
Table~\ref{tab_modpar} are the temperatures, as given by the gravity darkening
relation \citep{1963ApJ...138.1134C}, at the stellar equator and pole,
respectively. The plane-parallel input atmospheres are interpolated to these
values.

Again the line of {C}{\sc ii}$\lambda$\,4267 is most suited for constraining
$T_{\rm eff}$, at least when relying on NLTE line modeling, because its
strength strongly depends on $T_{\rm eff}$, while the modeled values of $i$ or
$M_\star$ turn out to hardly have an effect on either strength nor shape. We
note that systematic uncertainties of the behaviour of {C}{\sc
  ii}$\lambda$\,4267 may affect the value, however, within the frame of using
the SYNSPEC input model spectra, as computed above, the value is well
constrained. For this step, solar abundance values were assumed. The best fit
with a Bruce3 compute SED is obtained for $T_{\rm eff}=18\,500$\,K on the UVES
mean spectrum (see Fig.~\ref{fig_vsini}), and $T_{\rm eff}=18\,250$\,K on the
FORS1 mean spectrum. The fits for other spectral lines are reasonable as well,
but since those do vary more strongly with $i$ and $M_\star$ than {C}{\sc
  ii}$\lambda$\,4267 they have less power in constraining $T_{\rm eff}$.

An alternative, independent way to constrain $T_{\rm eff}$ is to use the
International Ultraviolet Explorer \citep[IUE,][]{1978Natur.275..372B}
flux-calibrated spectra and visual photometry, de-redden them with an
appropriate extinction law, and then fit the model fluxes. This not only gives
the temperature by fitting the slope of the UV spectrum, but also the distance
by fitting the absolute flux as measured above the Earth's atmosphere vs.\ the
absolute flux as modeled at a unit distance from the star. The latter will be
used in a later step, for now we perform the exercise only on the slope to
obtain $T_{\rm eff}$. Two spectra exist, both taken in high resolution with
the short wavelength arm (SWP camera). As for both the large aperture was
used, they can be considered as spectrophotometric, and are, in fact, in good
agreement with each other. Both spectra were obtained from the IUE archive,
co-added and finally median filtered to reduce the noise. For the photometry
we used the Str\"omgren values given by \citet{1976A&AS...25..213G} and the
flux calibration of \citet{1998AJ....116..482G}.

The value for the color excess $E_{(B-V)}=0.08$\,mag was taken from
\citet{1992AcA....42..211P}. However, they adopt a standard B2 star as
template, having an intrinsic color of $(B-V)_0=-0.21$, which due to the
He-strong nature of the object is not appropriate. There is consensus that
HR\,7355 is in fact cooler than a typical B2 star, namely about 17 to 18\,kK
\citep{2010A&A...511L...7M,oksala}, and thus an intrinsic color of
$(B-V)_0=-0.18$ might be a better estimate, so that the excess needs to be
adjusted to a lower value of $E_{(B-V)} = 0.05$\,mag. The effect of the
magnetic field on the colors at the temperatures, metallicities, and field
strenghts in question is entirely neglibible \citep{2005A&A...433..671K}.

To actually de-redden the IUE data, we use the extinction law in the
parametric form given by \citet{1989ApJ...345..245C}, assuming a standard
Galactic value of $R_V \equiv E_{(B-V)}/A_V=3.1$. An independent determination
of $R_V$ would require spectrophotometry of the 2200\,\AA\ extinction bump,
but unfortunately no IUE long-wavelength data were taken.  We note, however,
that changing $R_V$ does not modify the conclusions significantly, and in
particular does not provide any help in the distance problem presented below.
We find that the slope of the UV data between 1250 and 2000\,\AA\ is fitted
best with $T_{\rm eff}\sim17\,000$\,K. Since this seems a bit low compared to
the line determinations, we tried to fit a model for $T_{\rm eff}=18\,000$\,K
as well, but found this is only possible if we adopt a higher reddening of
$E_{(B-V)}=0.08$\,mag, which is argued above to be implausible. In addition,
it should be kept in mind that the star is chemically peculiar, after all, so
that an analysis based on the assumption of solar abundance might show some
systematic effects.

Since the measurements differ of $T_{\rm eff}$ somewhat, at this point we go
ahead with both options $T_{\rm eff}=17\,000$\,K and $T_{\rm eff}=18\,000$\,K,
and will decide at the end of the procedure which to adopt for the final
model.

Although the cool temperature derived here seems incompatible with the
spectral classification of B2, usually assumed to be at about 22\,000\,K, it
was mentioned above that the He-strong nature of the star is responsible for
the discrepancy. Early B stars are mainly classified on account of their
He{\sc i} lines, and since the helium lines are much stronger than they would
be for a star with solar abundances, classifications of B-type He-strong stars
are systematically biased towards the spectral type with the strongest He{\sc
  i} lines, which is B2.

%%%%%%%%%%%%%%%%%%%%%%%%%%%%%%%%%%%%%%%%%%%%%%%%%%%%%%%%%%%%%%%%%%%%%%%
%%%%%%%%%%%%%%%%%%%%%%%%%%%%%%%%%%%%%%%%%%%%%%%%%%%%%%%%%%%%%%%%%%%%%%%
%%%%%%%%%%%%%%%%%%%%%%%%%%%%%%%%%%%%%%%%%%%%%%%%%%%%%%%%%%%%%%%%%%%%%%%
\begin{table}
\caption[]{\label{tab_modpar}Stellar parameters. Only errors given in the
  ``input'' part indicate goodness of fit as discussed in the text, errors in
  the ``derived'' part are strictly the intervals derived from propagating the
  errors of the input parameters. Since the uncertainty of the period is
  insignificant wrt.\ that of the other quantities we treat it as fixed.}
\begin{center}
\begin{tabular}{lr}
\multicolumn{2}{c}{Bruce3 input parameters}\\
Stellar mass & $6\pm0.5$\,M$_{\sun}$\\
Inclination & $60\pm10\degr$ \\
Effective temperature &$17\,500\pm1000$\,K \\
$v\sin i$ & $310\pm5$\,km\,s$^{-1}$\\
$P_{\rm rot}$ & 0.5214404\,d\\
\\
\multicolumn{2}{c}{Bruce3 derived parameters}\\
$v_{\rm eq}$ & $358$\,km\,s$^{-1} \pm10\%$\\
$w=v_{\rm eq}/v_{\rm crit}$ & $0.89\pm7\%$\\
Polar radius & $3.06$\,R$_{\sun} \pm5\%$\\
Polar temperature & $19\,751$\,K $\pm7\%$ \\
Polar gravity (log) & $4.25 \pm11\%$ \\
Equatorial radius & $3.69$\,R$_{\sun} \pm10\%$\\
Equatorial temperature & $15\,740$\,K $\pm9\%$ \\
Equatorial gravity (log) & $3.85 \pm13\%$ \\
Luminosity & $972$\,L$_{\sun} \pm28\%$\\
\multicolumn{2}{c}{Adopted spectrophotometric parameters}\\
Excess $E_{(B-V)}$ & 0.065\,mag \\
Distance       & 236\,pc \\
\end{tabular}
\end{center}
\end{table}
%%%%%%%%%%%%%%%%%%%%%%%%%%%%%%%%%%%%%%%%%%%%%%%%%%%%%%%%%%%%%%%%%%%%%%%
%%%%%%%%%%%%%%%%%%%%%%%%%%%%%%%%%%%%%%%%%%%%%%%%%%%%%%%%%%%%%%%%%%%%%%%
%%%%%%%%%%%%%%%%%%%%%%%%%%%%%%%%%%%%%%%%%%%%%%%%%%%%%%%%%%%%%%%%%%%%%%%

%%%%%%%%%%%%%%%%%%%%%%%%%%%%%%%%%%%%%%%%%%%%%%%%%%%%%%%%%%%%%%%%%%%%%%%
%%%%%%%%%%%%%%%%%%%%%%%%%%%%%%%%%%%%%%%%%%%%%%%%%%%%%%%%%%%%%%%%%%%%%%%
%%%%%%%%%%%%%%%%%%%%%%%%%%%%%%%%%%%%%%%%%%%%%%%%%%%%%%%%%%%%%%%%%%%%%%%
\begin{figure}
\includegraphics[viewport=0 105 546 746,angle=0,width=9cm,clip=]{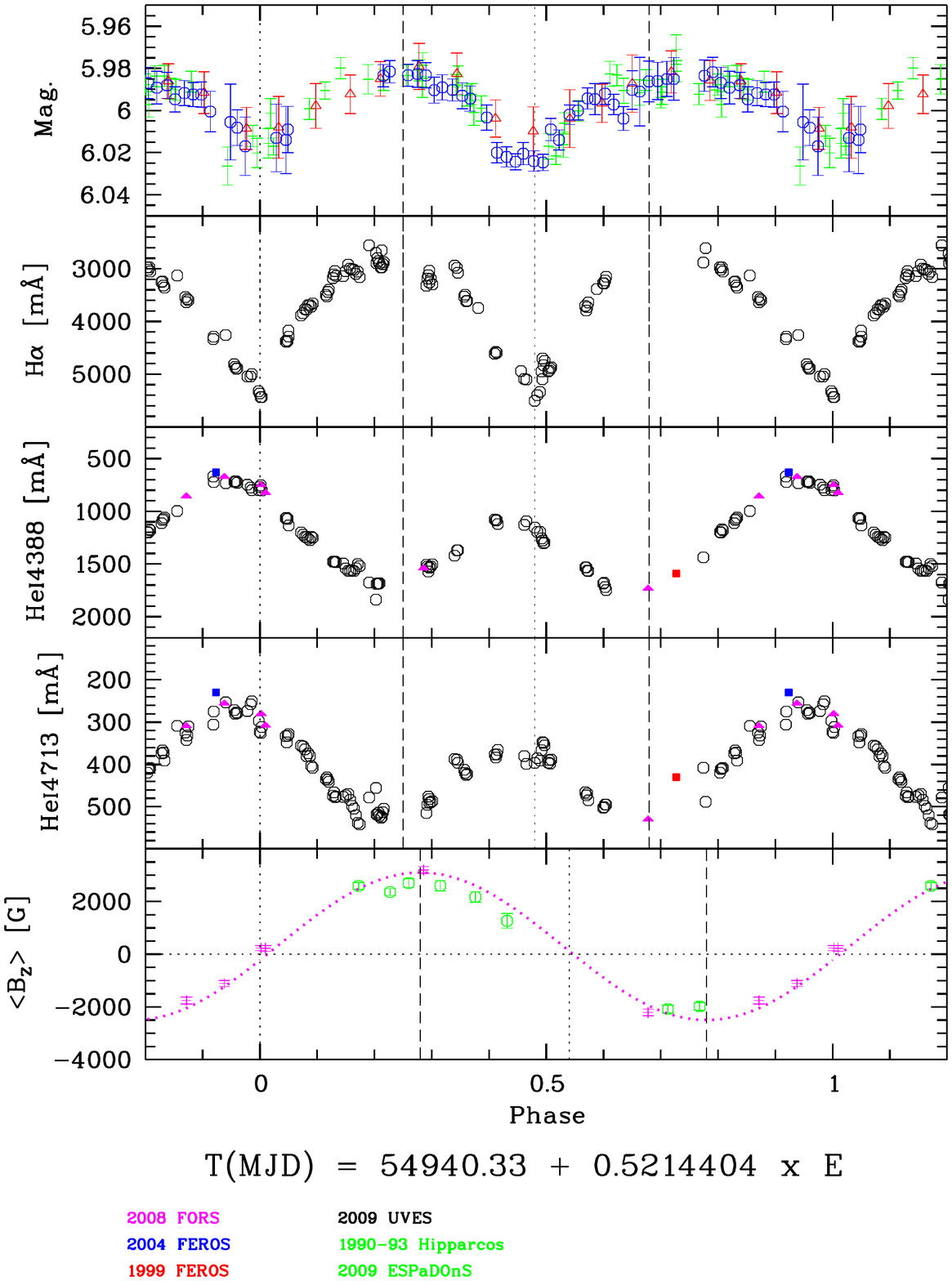}
\caption[]{Phased observational data.  Top panel: Photometric magnitudes from
  Hipparcos (green), \citet[][binned to 64 phase bins, blue open
    circles]{oksala}, and ASAS data (red triangles, binned to 16 phase bins)
  from \citet{2010A&A...511L...7M}.  Second, third, and fourth panels:
  Equivalent width measurements of H$\alpha$, He{\sc
    i}$\lambda$$\lambda$\,4388, and 4713 in m\AA, in the UVES data (black
  circles), the 2008 FORS1 spectra \citep[][purple triangles]{rivi7355}, and
  the 1999 and 2004 FEROS spectra \citep[][red and blue squares,
    respectively]{2008A&A...482..255R}.  Lowermost panel: Magnetic data from
  FORS1 \citep[][see there a version of this plot with less data]{rivi7355},
  and ESPaDOns magnetic data (open circles) in gauss and a sinusoidal fit to
  the FORS1 measurements. The vertical lines denote special phases discussed
  in the text.}
\label{fig_phasingall}
\end{figure}
%%%%%%%%%%%%%%%%%%%%%%%%%%%%%%%%%%%%%%%%%%%%%%%%%%%%%%%%%%%%%%%%%%%%%%%
%%%%%%%%%%%%%%%%%%%%%%%%%%%%%%%%%%%%%%%%%%%%%%%%%%%%%%%%%%%%%%%%%%%%%%%
%%%%%%%%%%%%%%%%%%%%%%%%%%%%%%%%%%%%%%%%%%%%%%%%%%%%%%%%%%%%%%%%%%%%%%%

\subsection{Inclination and stellar mass}

In terms of Bruce3 modeling, since all other relevant quantities, like period
and $v \sin i$, are well constrained, the inclination effectively sets the
model's equatorial radius (since it converts $v\sin i$ to the rotation
velocity) and the geometric projection onto the line of sight.  As the
equatorial radius is thus constrained, the modeling parameter range is chosen
to include masses between 6 and 8\,M$_{\sun}$, considering stellar evolution:
lower mass stars at the main sequence are not large enough for the required
equatorial radius, neither are higher mass stars small enough. The considered
range is sufficiently broad to to rely on details of the evoolution, it is,
after all, just the range over which a grid is computed. The possible inferred
range of effective temperature supports this mass range.

In terms of modeling, the stellar mass only scales the polar radius and the
amount of gravity darkening. However, both quantities have only subtle effects
on the modeled spectral lines, and results are not really conclusive.
However, in one respect the model does vary strongly with these two
parameters, namely in the total luminosity and in the flux seen by the observer,
since they scale the size of the stellar surface through the radius and the
surface temperature distribution. Therefore, the traditional approach,
determining the distance with the modeled flux, can be inverted: The revised
Hipparcos parallax is sufficiently precise to do so. The Hipparcos distance,
$273\pm26$\,pc, is used to define a range for model fluxes; namely ``near''
(247\,pc), ``mean'' (273\,pc), and ``far'' (299\,pc) distances.

We find the models for $i=80\degr$ do not have sufficient flux; they remain
well under the faintest flux limit, even for the ``near'' distance.  For
$T_{\rm eff}=18\,000$\,K, models for $i=70\degr$ are just in agreement with
the ``near'' distance flux, and only models with $i=60\degr$ are compatible
with the flux curve for the ``mean'' distance. For $T_{\rm eff}=17\,000$\,K,
even models with $i=60\degr$ are just compatible with the flux curve for the
``near'' distance. The ``far'' flux is too luminous to be reproduced by any of
the computed parameter sets. The ultimately-adopted inclination is, therefore,
$i=60\degr$ for which we assume a $\pm10\degr$ uncertainty.

Although this still does not constrain the mass in a straightforward
observational way, all the other parameters are sufficiently well-known to
leave only a narrow range of physically acceptable masses. This is because we
can, at least to first order, expect that the luminosity of a star of a
given mass does not change with its rotational velocity.

The luminosity computed by Bruce3 modeling, however, is not constrained by the
physical processes inside the stellar interior; it is just given by the
temperature distribution and the stellar surface area, both of which can be
chosen independently in Bruce3. The stellar evolution requirement that mass
and luminosity are related is an independent constraint, taken into account
only in a later step (and as well for setting a reasonable grid range
above). This means the Bruce3-computed luminosity is almost invariant with
model mass. Instead, it increases with decreasing inclination, in such a way
that all $T_{\rm eff}=18\,000$\,K, $i=60\degr$ models have a luminosity of
about $L_\star = 1120\,{\rm L_{\sun}}$, $T_{\rm eff}=17\,000$\,K, $i=60\degr$
models have a luminosity of about $L_\star = 1000\,{\rm L_{\sun}}$. Such a
luminosity is only compatible with a mass of about
$6\pm0.5$\,M$_{\sun}$. Unfortunately, both models give a very similar distance
to match the absolute flux, around 240\,pc, which is only somewhat closer than
the Hipparcos distance of $273\pm26$\,pc.

%%%%%%%%%%%%%%%%%%%%%%%%%%%%%%%%%%%%%%%%%%%%%%%%%%%%%%%%%%%%%%%%%%%%%%%
%%%%%%%%%%%%%%%%%%%%%%%%%%%%%%%%%%%%%%%%%%%%%%%%%%%%%%%%%%%%%%%%%%%%%%%
%%%%%%%%%%%%%%%%%%%%%%%%%%%%%%%%%%%%%%%%%%%%%%%%%%%%%%%%%%%%%%%%%%%%%%%
\begin{figure*}
\parbox{0.4cm}{~}%
\parbox{4.4cm}{\centerline{[He{\sc i}]$\lambda$\,4045}}%
\parbox{4.4cm}{\centerline{He{\sc i}$\lambda$\,4713}}%
\parbox{4.4cm}{\centerline{He{\sc i}$\lambda$\,6678}}%
\parbox{4.4cm}{\centerline{Ne{\sc i}$\lambda$\,6717}}%

\includegraphics[viewport=100 188 608 1010,angle=0,width=4.4cm,clip=]{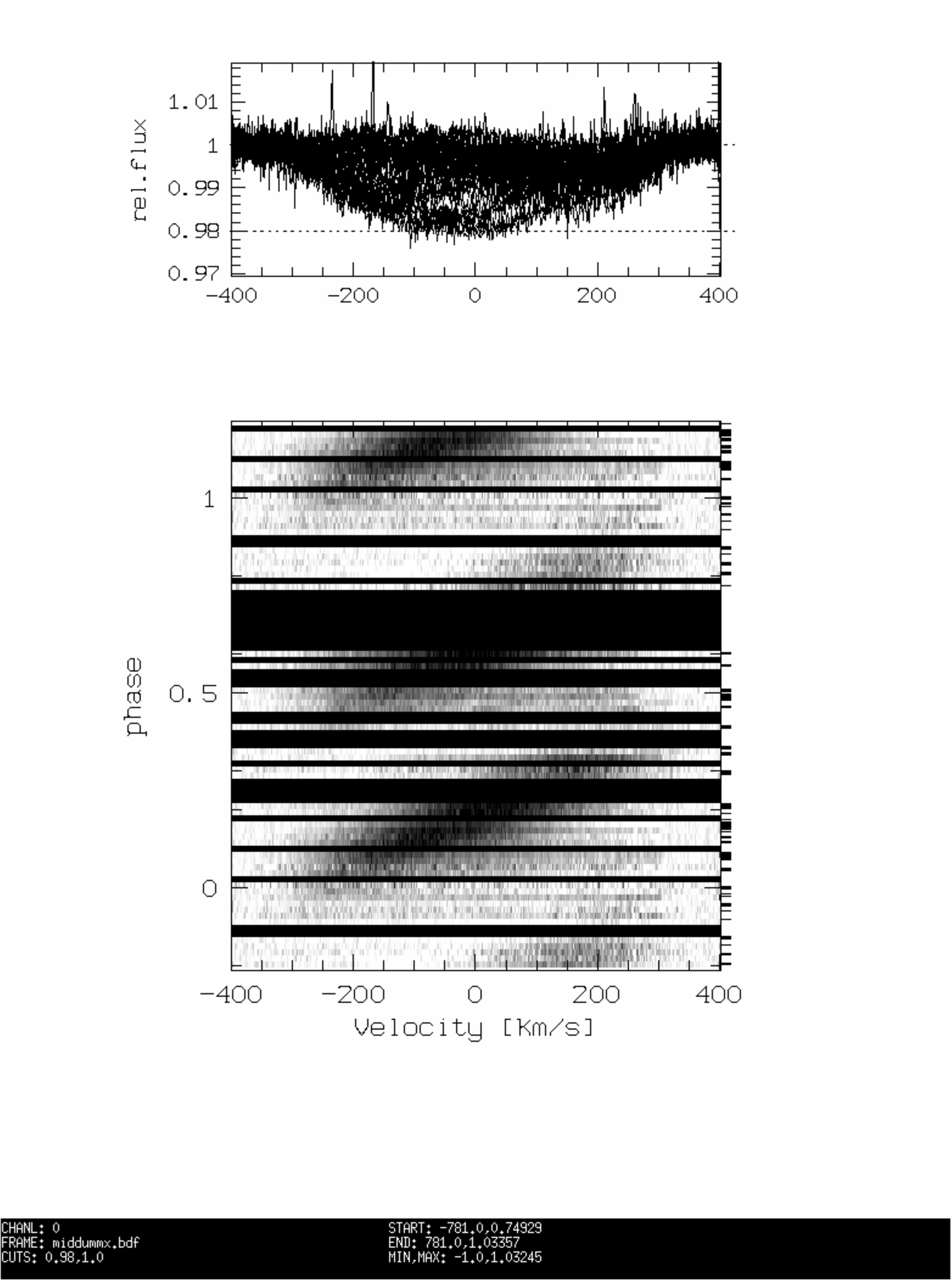}%
\includegraphics[viewport=100 188 608 1010,angle=0,width=4.4cm,clip=]{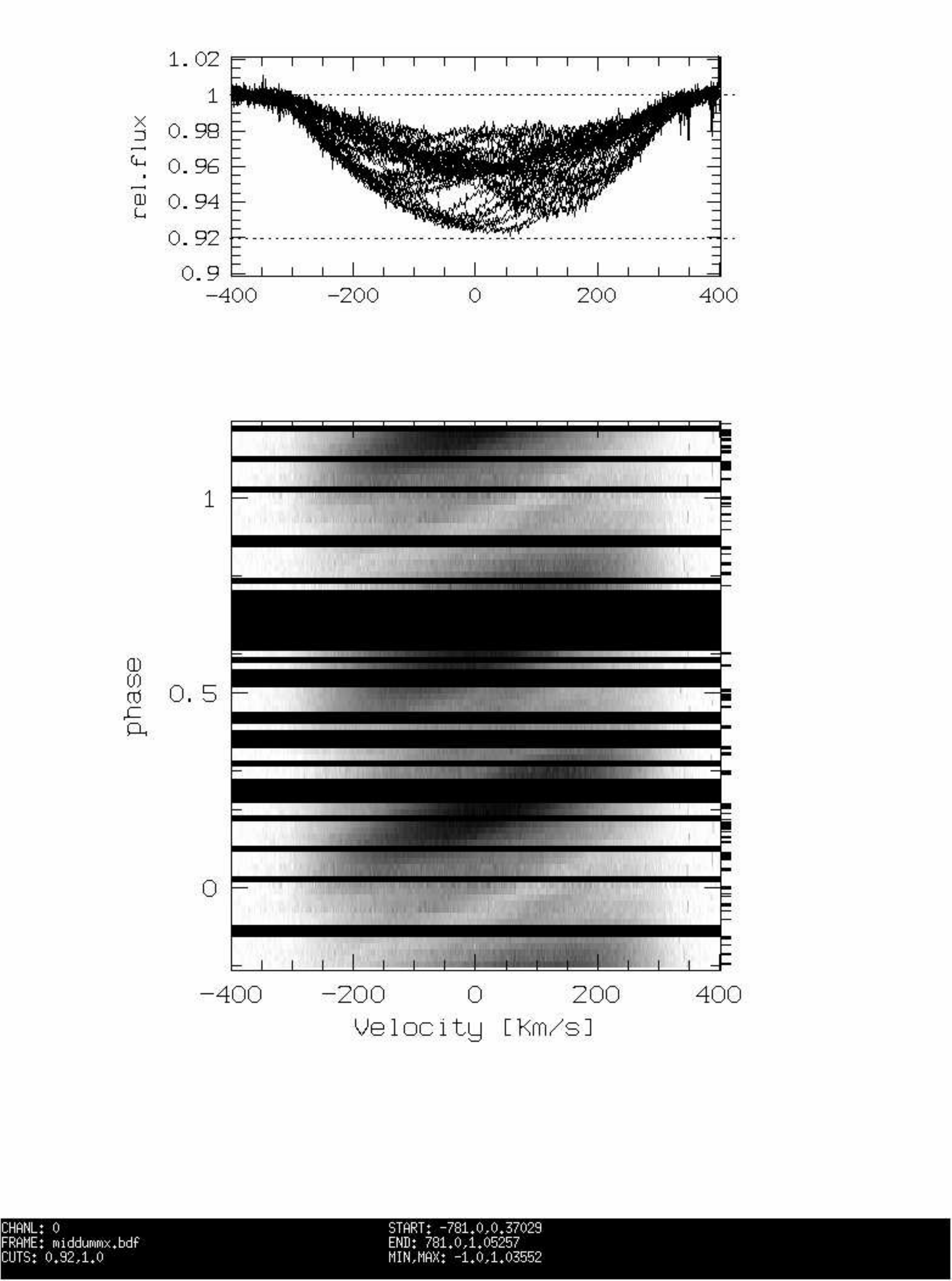}%
\includegraphics[viewport=100 188 608 1010,angle=0,width=4.4cm,clip=]{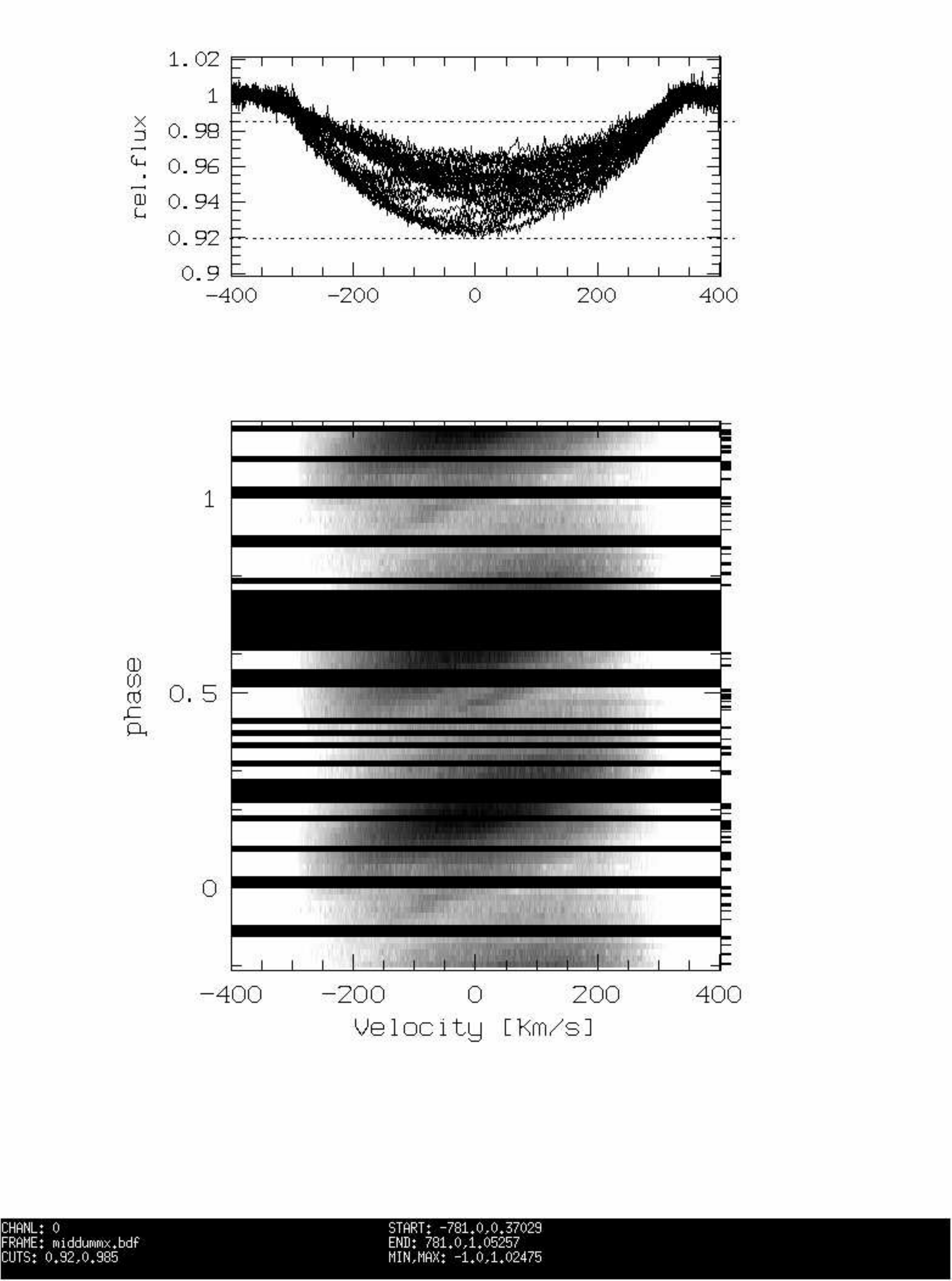}%
\includegraphics[viewport=100 188 608 1010,angle=0,width=4.4cm,clip=]{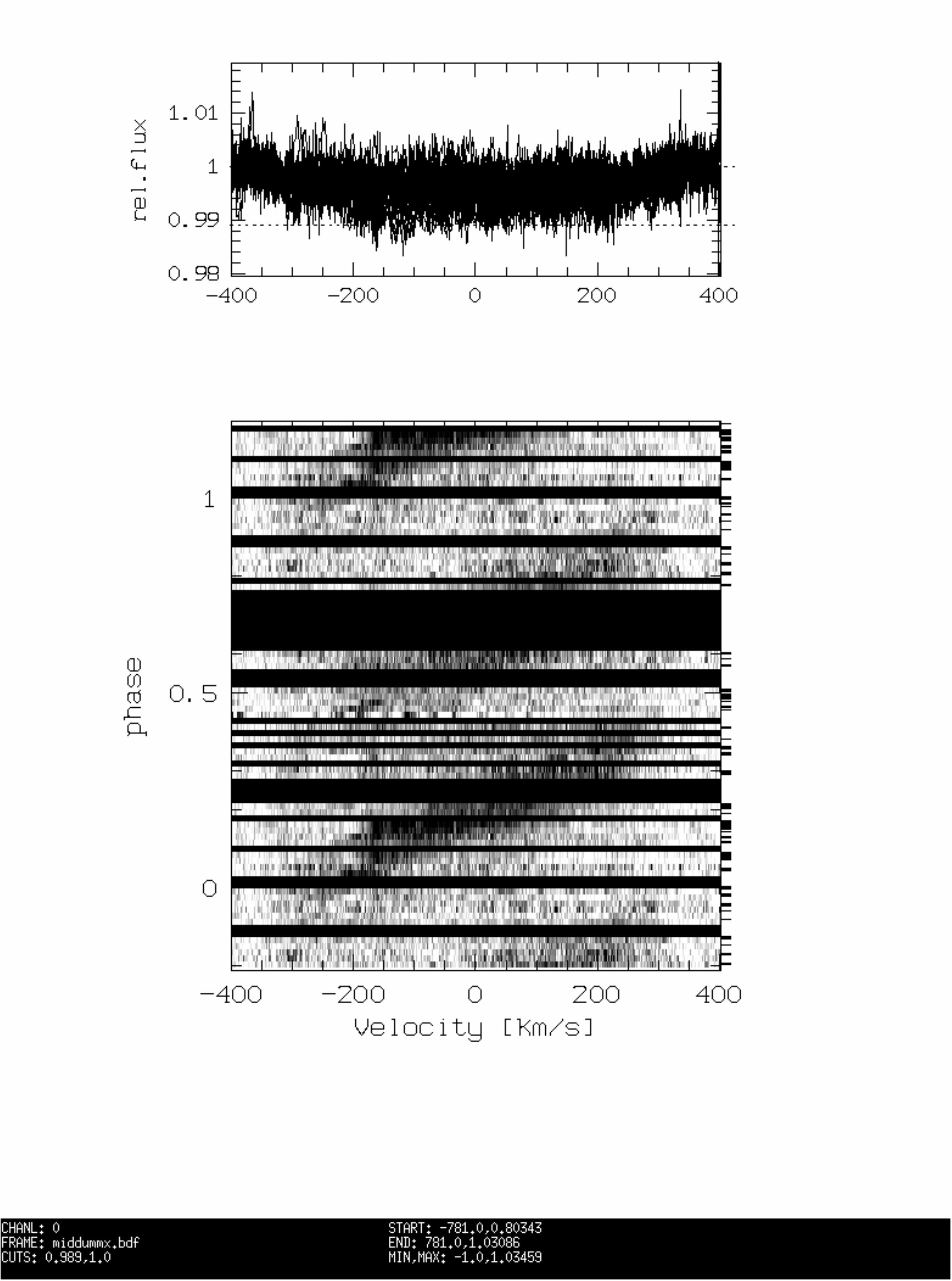}%

\parbox{0.4cm}{~}%
\parbox{4.4cm}{\centerline{N{\sc ii}$\lambda$\,4631}}%
\parbox{4.4cm}{\centerline{C{\sc ii}$\lambda$\,4267}}%
\parbox{4.4cm}{\centerline{Si{\sc iii}$\lambda$\,4553}}%
\parbox{4.4cm}{\centerline{S{\sc ii}$\lambda$\,4816}}%

\includegraphics[viewport=100 188 608 1010,angle=0,width=4.4cm,clip=]{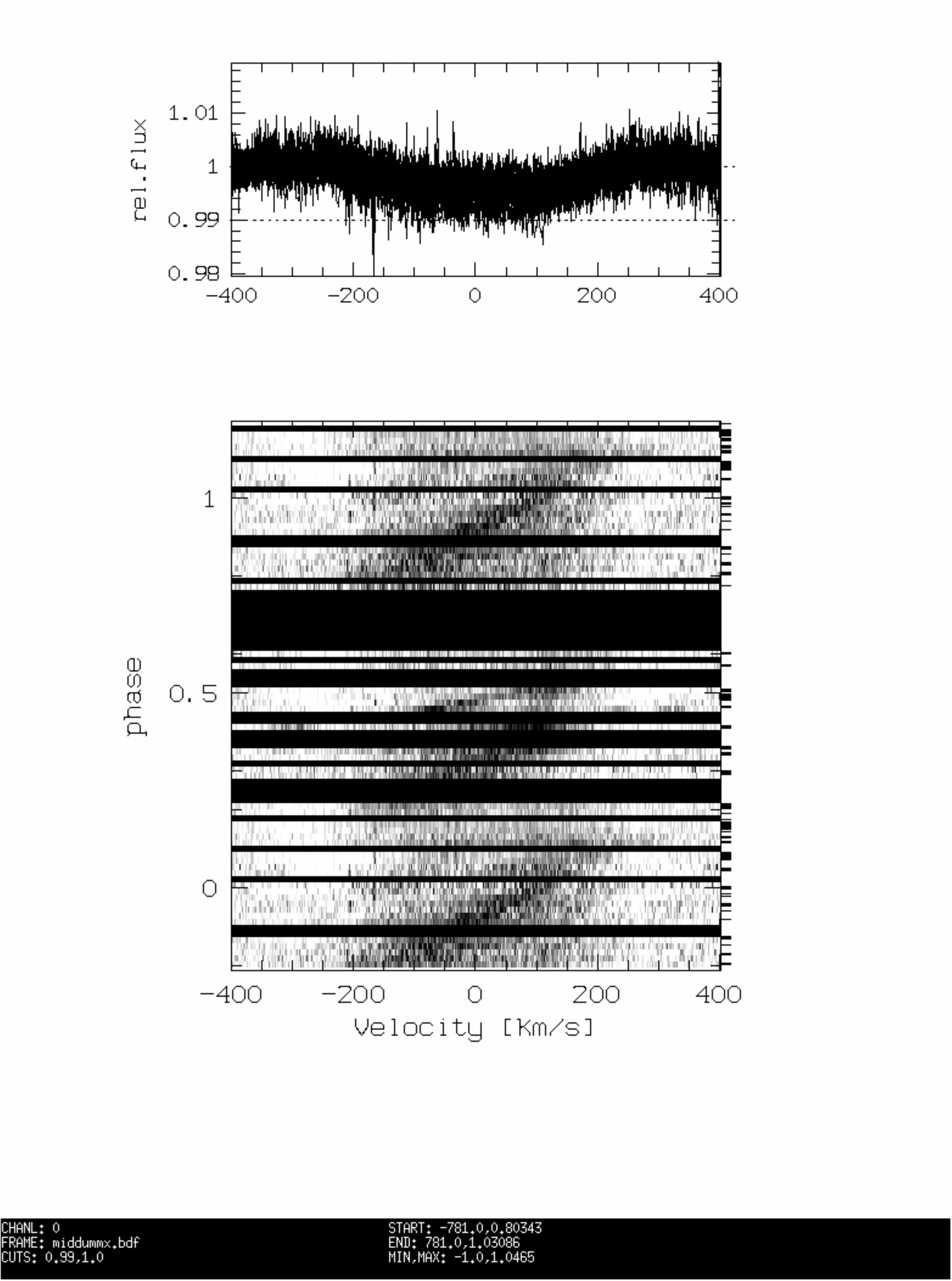}%
\includegraphics[viewport=100 188 608 1010,angle=0,width=4.4cm,clip=]{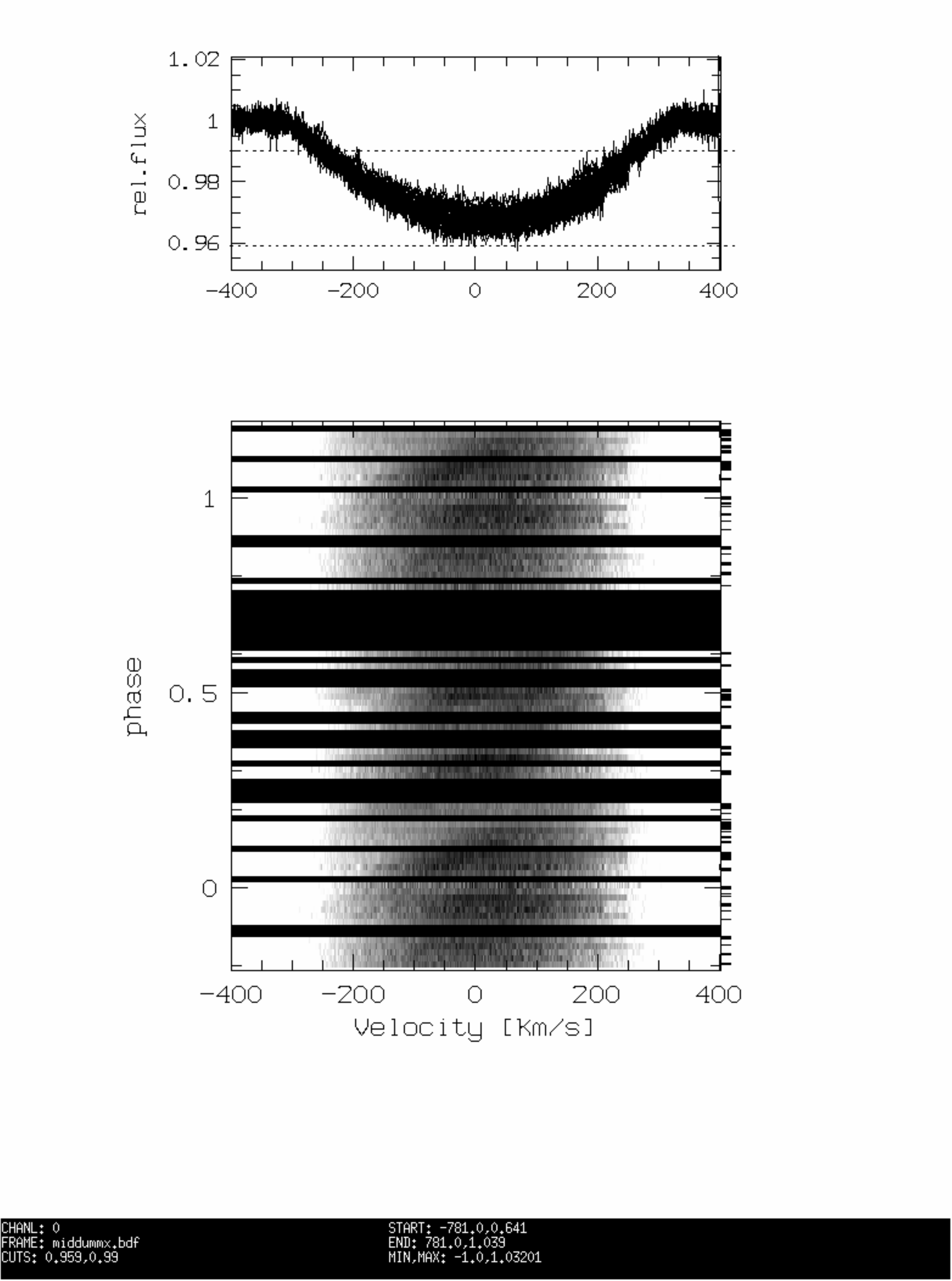}%
\includegraphics[viewport=100 188 608 1010,angle=0,width=4.4cm,clip=]{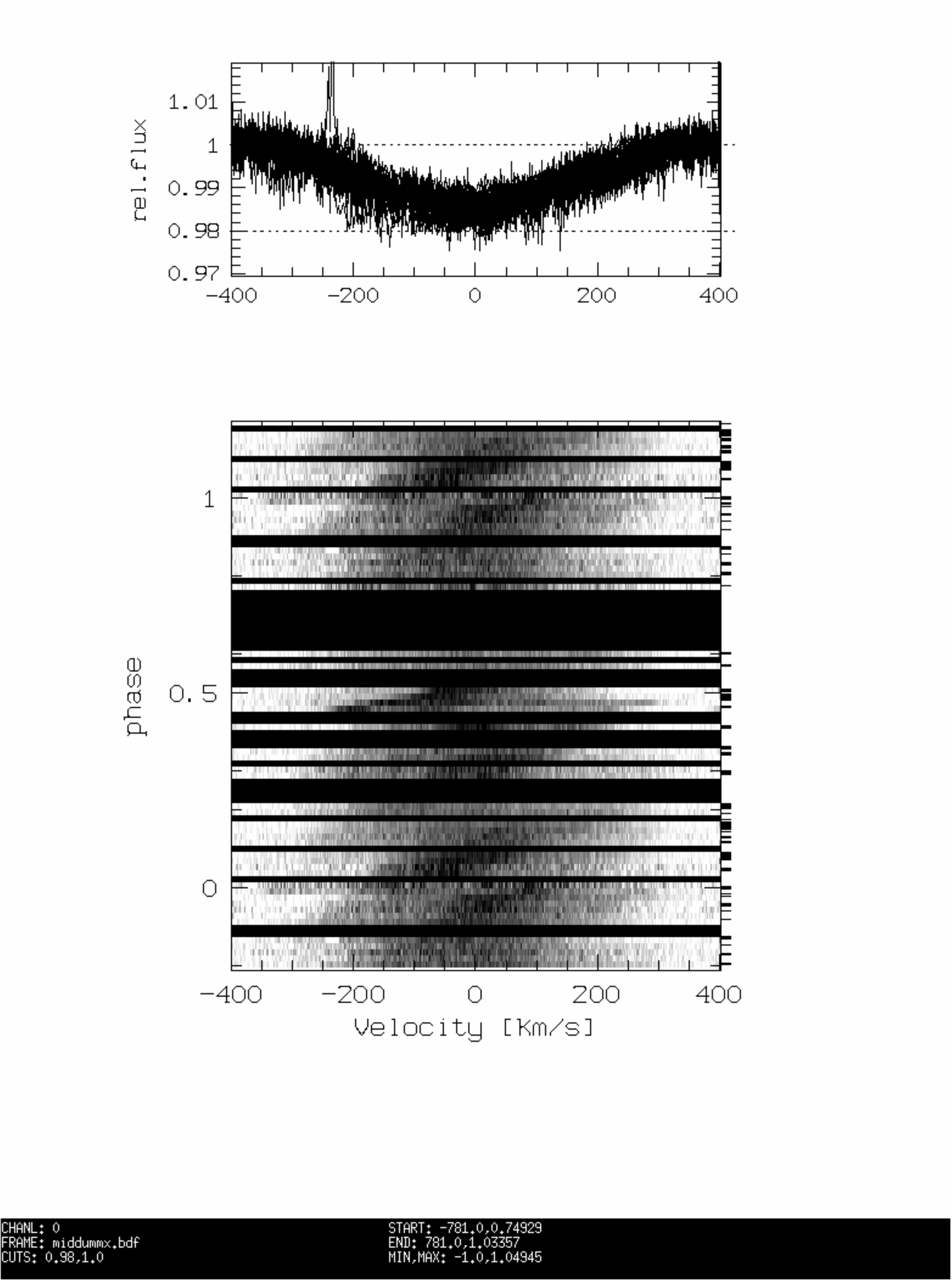}%
\includegraphics[viewport=100 188 608 1010,angle=0,width=4.4cm,clip=]{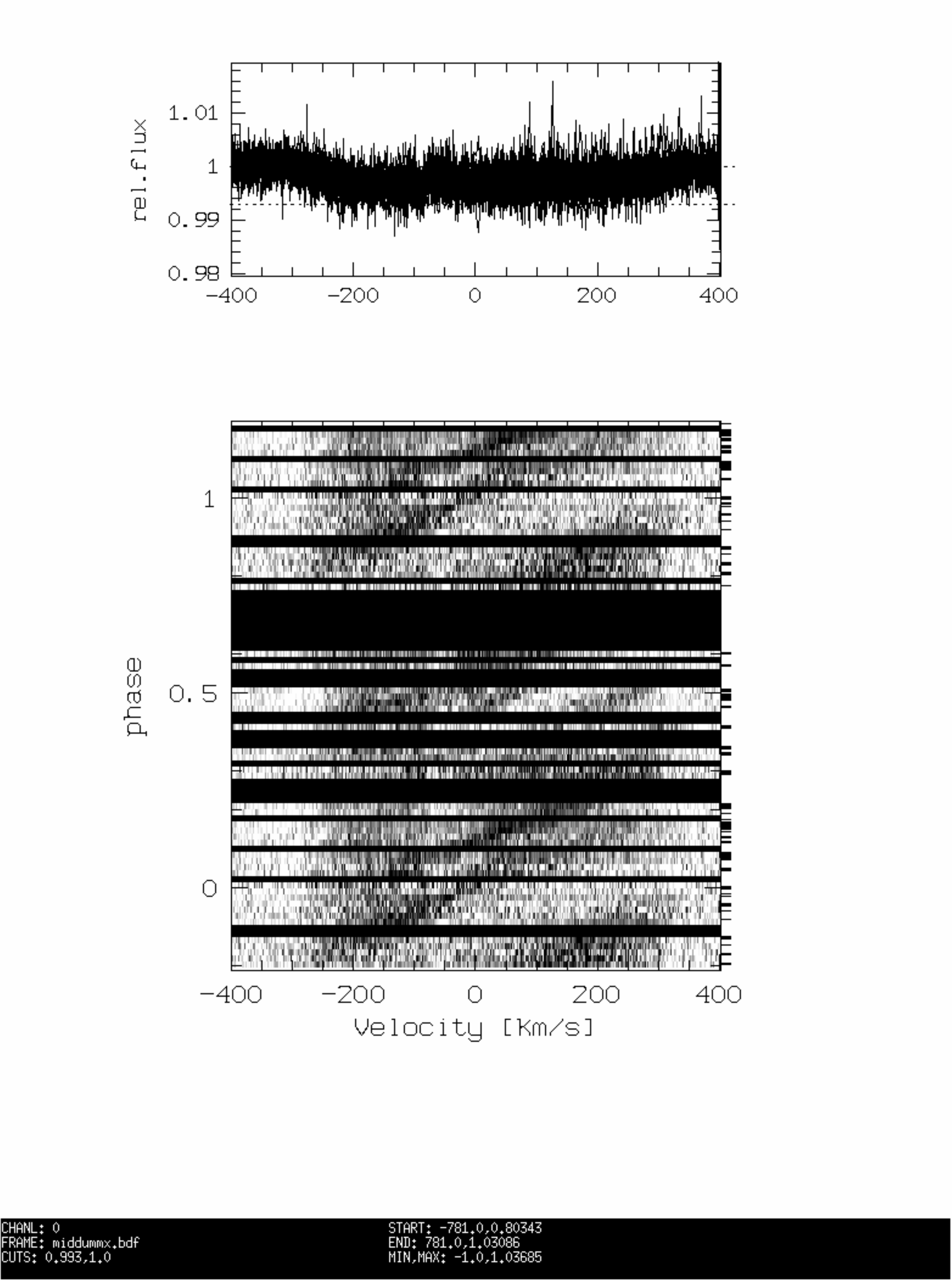}%

\caption[]{Phased variations of photospheric helium and metal lines,
  represented as dynamic spectra. Top panels: original profiles, the dotted
  lines mark the cuts applied for black and white in the grey-scale
  panels. Bottom panels: phased dynamical spectra.  These panels were
  constructed with 32 or 64 phase bins (64: Si{\sc iii}$\lambda$\,4553 and
  S{\sc ii}$\lambda$\,4816, He{\sc i}$\lambda$\,6678, He{\sc
    i}$\lambda$\,4713) depending on the complexity of the variations prompting
  for a trade-off of higher sampling vs.\ the noise level. For a detailed
  description of the variations see Sect.~\ref{sec_photo}.}
\label{fig_dynamicalphoto}
\end{figure*}
%%%%%%%%%%%%%%%%%%%%%%%%%%%%%%%%%%%%%%%%%%%%%%%%%%%%%%%%%%%%%%%%%%%%%%%
%%%%%%%%%%%%%%%%%%%%%%%%%%%%%%%%%%%%%%%%%%%%%%%%%%%%%%%%%%%%%%%%%%%%%%%
%%%%%%%%%%%%%%%%%%%%%%%%%%%%%%%%%%%%%%%%%%%%%%%%%%%%%%%%%%%%%%%%%%%%%%%

So the only remaining question, left open above, is whether to adopt 17\,000
or 18\,000\,K. While the model for 18\,000K agrees better with the
spectroscopic results, and has a more plausible luminosity, it would require a
somewhat higher reddening than suggested by $E_{(B-V)}$.  On the other hand,
the model for 17\,000\,K provides a better fit for the spectral energy
distribution, yet its luminosity would be lower by 10 to 20\,\% than expected
even for a star of 6\,M$_{\sun}$ only. However, given the relatively low
extinction this is a rather robust determination. For the final model,
therefore, we adopt 17\,500\,K, but unfortunately have to accept a rather
generous error margin of 1000\,K.

The final parameters are summarized in Table~\ref{tab_modpar}, the absolute
flux fit for these parameters is shown in Fig.~\ref{fig_fluxfit}. The figure
illustrates that these parameters give good fits to both absolute flux and
line profiles, except of course the variable helium lines, as well as some of
the UV lines.  Concerning the uncertainty intervals given in
Table~\ref{tab_modpar}, we note that only the one given in the uppermost part
(input parameters) were constrained traditionally, i.e.\ by judging a grid of
model vs.\ observational data. The intervals given for the derived parameters,
however, were obtained by propagating these input parameter uncertainties
through the model code, and thus they do not necessarily satisfy the usual statistical
relations (in particular, they are non-orthogonal to each other). Furthermore,
since the quantities have been computed, we give more significant digits than
the interval would warrant, to keep the values of the derived parameters
consistent with the input parameters.

%%%%%%%%%%%%%%%%%%%%%%%%%%%%%%%%%%%%%%%%%%%%%%%%%%%%%%%%%%%%%%%%%%%%%%%
\section{Photospheric abundance pattern}
\label{sec_photo}

Figure~\ref{fig_phasingall} shows previously published data and equivalent
widths measured in the UVES data, phased with the ephemeris given in
Sect.~\ref{sec_ephem}. The uppermost panel shows the available photometric
data.  A constant offset on magnitude has been added to the additional data
sets, taken in different photometric systems, for the purpose of display
together with the Hipparcos data. We note that photometric data from different
bands may not have the same amplitude, as well as shifts in phase (expressed
here a linear phase $\phi$) vs.\ each other, but the accuracy of the
available data is not sufficient to determine if this is the case here.

In any case, the data by \citet{oksala} and the UVES spectra were taken only
one year apart, and the period is known well enough to be sure now about the
phasing of the spectroscopic variations vs.\ the photometric: When H$\alpha$
is in occultation, there is a photometric minimum. However, since the
photometric minima are broad and the phases of occultations and He{\sc i} EW
minima differ only by $\Delta\phi\approx 0.07$, the following can also be
said: When the He{\sc i} absorption is weakest, there is a photometric
minimum.  Whether the photometric variations are due to the occultations, as
suggested by \citet{2008MNRAS.389..559T}, or due to the photospheric abundance
pattern, as suggested by \citet{2010A&A...511L...7M}, will require careful
modeling, and possibly further multicolor photometric observations to
distinguish geometric from intrinsic effects.

%%%%%%%%%%%%%%%%%%%%%%%%%%%%%%%%%%%%%%%%%%%%%%%%%%%%%%%%%%%%%%%%%%%%%%%
\subsection{Line strength variations}
The H$\alpha$ equivalent widths are fully dominated by variations of the
circumstellar environment; there is no indication of an intrinsic photospheric
variation of this or any other hydrogen line strength (see
Fig.~\ref{fig_dynamicalmagneto}). The vertical dotted lines in the four upper
panels of Fig.~\ref{fig_phasingall} mark the central points of the
occultations, i.e.\ when the circumstellar lobes are centrally in front of the
stellar disk (at $v_{\rm sys}$). This is at $\phi=0$, actually fixed by the
definition of the ephemeris, and at $\phi=0.46$.

The third and fourth panels in Fig.~\ref{fig_phasingall} show the equivalent
width variations of the He{\sc i}$\lambda$$\lambda$\,4713 and 4388 lines.  The double wave
character of the He{\sc i} equivalent width curve is very clear. Although the
strongest EW points in both half-cycles are of a similar level, the weakest EW
values differ by a factor of two. The dashed lines mark the estimated times of
maximal He-line strength at about $\phi=0.25$ and $0.68$.

%%%%%%%%%%%%%%%%%%%%%%%%%%%%%%%%%%%%%%%%%%%%%%%%%%%%%%%%%%%%%%%%%%%%%%%
\subsection{Line profile variations}

The most obvious photospheric variation is that of the He{\sc i} lines. The
variations are very strong, relative to the line strength much stronger than
in the prototype $\sigma$\,Ori~E: e.g.\ the EW of He{\sc i}$\lambda$\,4713
varies between 250 and 550\,m\AA\ in HR7355 vs.\ between 420 and 570\,m\AA\ in
$\sigma$\,Ori~E (measured in our own and archival data for both stars).
Figure~\ref{fig_phasingall} shows a factor of two and four between the EWs of
the weak and strong states of He{\sc i} lines. Judging the EW variations in
the figure at face value, they form a simple double wave pattern that would be
well compatible with just two enhanced polar regions and a depleted belt. A
detailed look at the spectra, however, shows that the variations are not that
easily explained.

The forbidden transition of He{\sc i} at
4045\,\AA\ (Fig.~\ref{fig_dynamicalphoto}, upper left) shows the simplest
pattern; presumably because it is the weakest line of He{\sc i} only the
strongest features are seen: Two He-enhanced regions are visible, which,
according to the phasing with the magnetic curve in Fig.~\ref{fig_phasingall},
are consistent with being located at about the magnetic poles, as is typical
for He-strong stars \citep[see e.g.][for $\sigma$\,Ori
  E]{1982A&A...116...64G,2000A&A...363..585R}.  These two regions are also the
dominant ones in other helium lines. They cross $v_{\rm
  sys}=v_0=7$\,km\,s$^{-1}$ at $\phi_0=0.19$ and $\phi_0=0.65$.

However, the stronger lines show additional features. In He{\sc
  i}$\lambda$\,4713 and 6678 (Fig.~\ref{fig_dynamicalphoto}, middle two panels
in upper row) three more, albeit weak, features are seen: They cross zero
velocity at $\phi_0=0.02$, $0.46$, and $0.88$. The position of the first two
agree with the occultations (see Sect.~\ref{sec_magnetosphere}). However, the
time these features take to propagate across the profile (formally an inverse
acceleration, in dynamical spectra directly visible as ``slope'' of a
propagating feature), show this is not the case: The time to cross the entire
range of 2 $v \sin i$ for undoubted occultation features, in all lines, is
much faster, as seen at $\phi_0=0.46$ in the panels for N{\sc
  ii}$\lambda$\,4631 and Si{\sc iii}$\lambda$\,4553 in
Fig.~\ref{fig_dynamicalphoto}. This is so because the magnetosphere is empty
(or at least far less dense) closer to the star than about 2 stellar radii, as
seen by the lack of H$\alpha$ emission at velocities $<2\times v \sin i$ at
quadrature, and the crossing time of a corotating feature scales inversely
with the distance from the star.

In the fairly many Ne{\sc i} lines detected in the spectrum, like Ne{\sc
  i}$\lambda$\,6717 shown in Fig.~\ref{fig_dynamicalphoto} (upper right panel), there
are two features crossing the profile, at $\phi_0=0.20$ and
$\phi_0=0.64$. These are almost the same phases as for the strong helium
features. Ne{\sc i}, therefore, is a metallic species that is probably
enriched at the magnetic poles, rather than at the magnetic equator, as one
usually observes for metals in He-strong stars.

Unfortunately, looking at more metal lines does not make the picture any
simpler: While in N{\sc ii}$\lambda$\,4631 only two features are seen ($\phi_0=0.35$
and $0.93$) four each can be identified in C{\sc ii}$\lambda$\,4267 and Si{\sc
  iii}\,4553 ($\phi_0=0.08, 0.32, 0.49$, and $0.95$), and possibly as many as
six are seen in S{\sc ii}$\lambda$\,4816 ($\phi_0=0.02, 0.13, 0.32, 0.49, 0.60$, and
$0.73$).

%%%%%%%%%%%%%%%%%%%%%%%%%%%%%%%%%%%%%%%%%%%%%%%%%%%%%%%%%%%%%%%%%%%%%%%
%%%%%%%%%%%%%%%%%%%%%%%%%%%%%%%%%%%%%%%%%%%%%%%%%%%%%%%%%%%%%%%%%%%%%%%
\subsection{Doppler imaging}\label{sec_DI}
We applied the Doppler imaging (DI) technique to interpret phase-dependent
behavior of some of the variable spectral lines in terms of an inhomogeneous
distribution of chemical elements. Due to the strong rotational Doppler
broadening of its spectra, HR\,7355 is a very challenging target for DI. In
fact, it has the largest $v\sin i$ among the early-type stars to which
abundance DI has been applied. Due to excessive blending and shallow line
profiles, only C{\sc ii}$\lambda$\,4267 and strong He lines, such as He{\sc
  i}$\lambda$\,4713, are suitable for modeling. We used these two spectral
features to derive respective abundance distributions with the updated version
of the magnetic DI code described by \citet{2002A&A...381..736P}.

Since HR\,7355 has a very strong magnetic field, it has to be taken into
account in abundance mapping. Available Stokes $I$ observations do not allow
the derivation of the magnetic field geometry simultaneously with chemical
spots. We therefore adopted the 11.6~kG dipolar magnetic field model derived
in the next section. On the other hand, gravity darkening and oblateness of
the star were not included in the modeling. As a consequence, we might expect
spurious pole-to-equator abundance gradients for the lines especially
sensitive to effective temperature. For the prupose of DI, stellar parameters,
i.e.\ inclination and $v \sin i$, were adopted as derived above form
spectroscopic modeling.

\begin{figure*}

\includegraphics[angle=0,width=12cm,clip=]{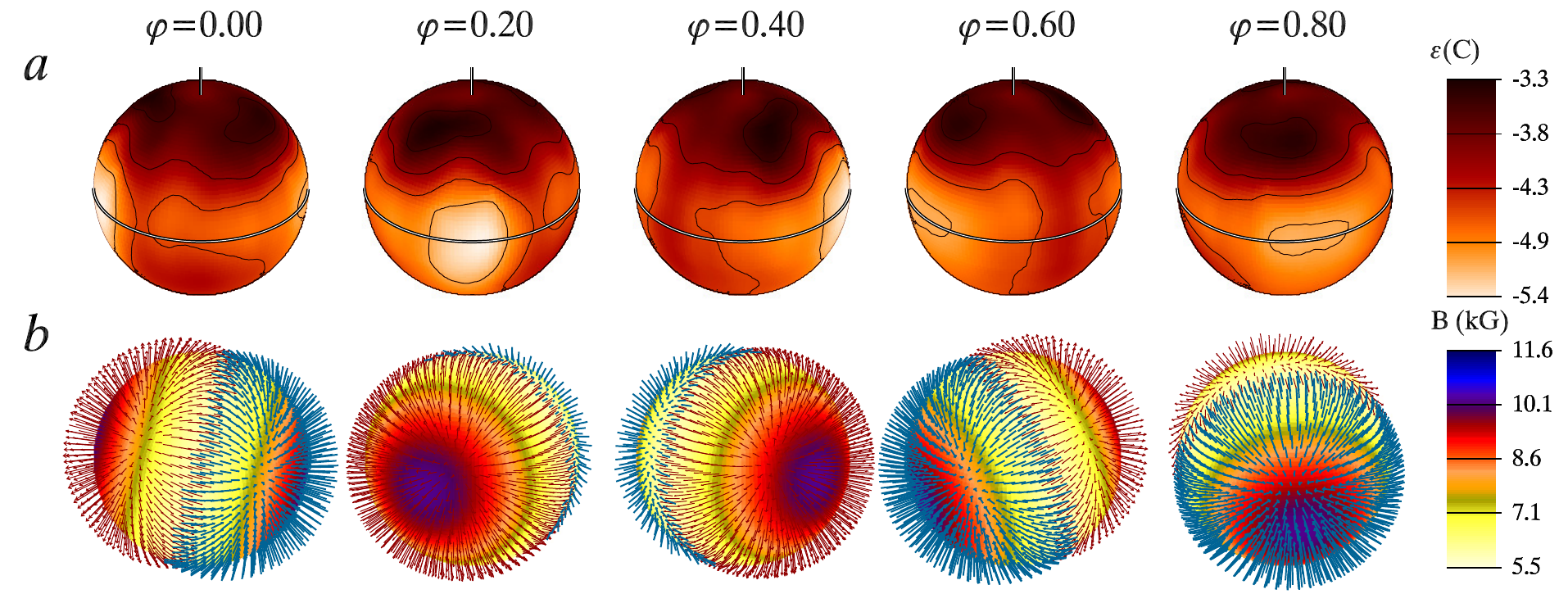}%

\includegraphics[angle=0,width=12cm,clip=]{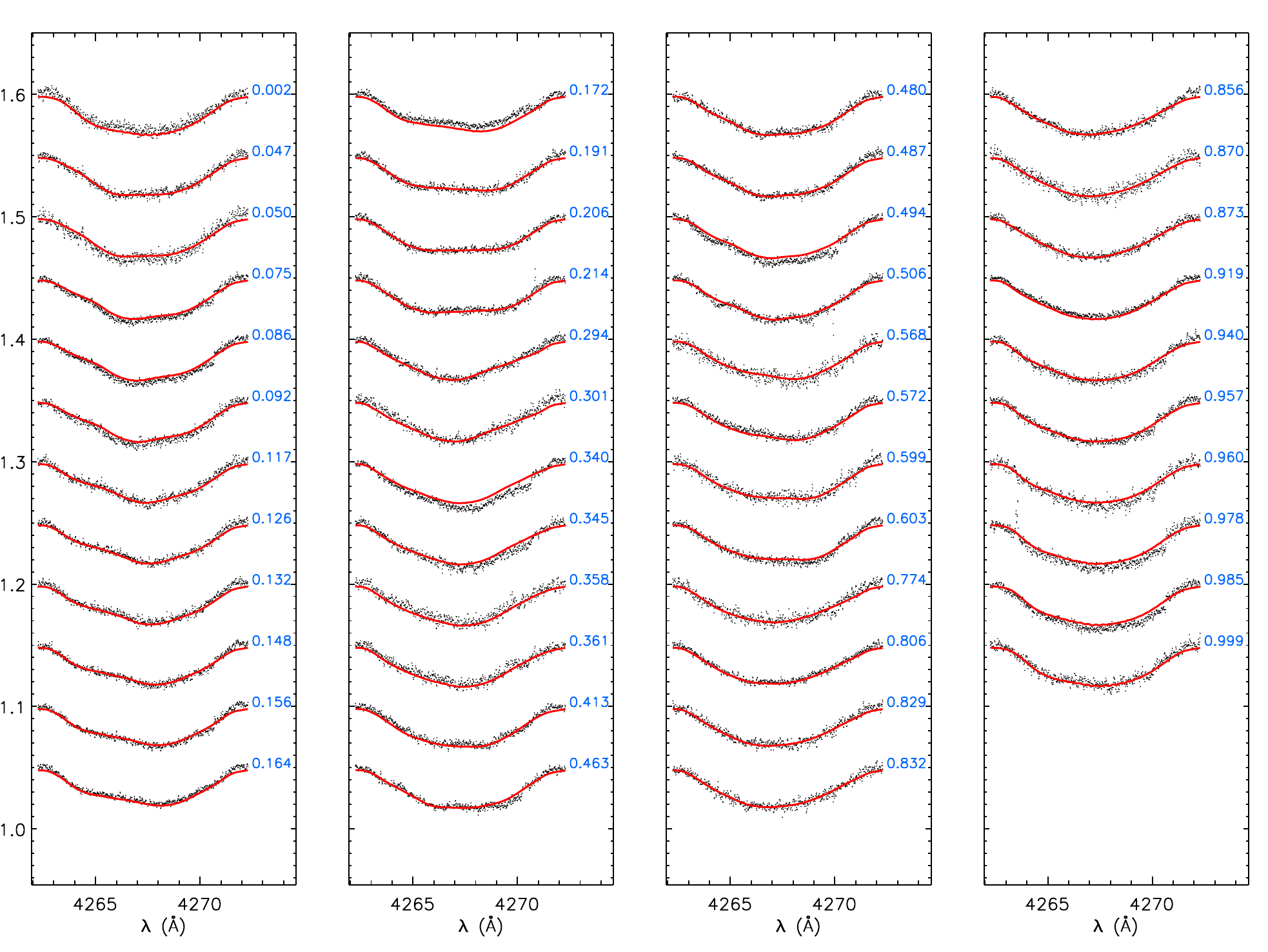}%

\caption[]{Carbon abundance distribution reconstructed from the variations of
  the C{\sc ii}$\lambda$\,4267 line. The top panels show spherical maps of the abundance
  distribution labelled by phase with iso-abundance lines plotted with a
  0.5~dex step (a) and assumed dipolar magnetic field geometry (b).  The C
  abundance is given in the $\log (N_{\rm C}/N_{\rm tot})$ units.}
\label{fig_DI_CII}
\end{figure*}

The results of the DI analysis of C{\sc ii}$\lambda$\,4267 are illustrated in
Fig.~\ref{fig_DI_CII}. The surface distribution of carbon required to
reproduce variability of this line features significant meridional gradient,
possibly due to unaccounted gravity darkening. In addition, there is about
1~dex abundance contrast at high latitudes, dominated by two spots seen at
rotational phases 0.25 and 0.80.

Helium is overabundant in the atmosphere of HR\,7355 and shows a high-contrast
surface distribution resulting in dramatic variability of He{\sc i}
lines. Since concentration of He is comparable to that of H over a significant
part of the stellar surface, not only is the observed line profile variability
a result of the abundance dependence of the He line absorption coefficient,
but is is also a consequence of continuum instensity variations and, possibly,
lateral variations of the atmospheric structure.  All these effects have to be
accounted for in order to obtain an accurate absolute scale and magnitude of
the He abundance variations. We carried out He mapping with the help of a new
magnetic DI code, {\sc Invers13} \citep{2012arXiv1201.1902K}, which uses a
grid of model atmospheres to treat self-consistently the impact of abundance
variations on the line profiles. Our present calculations are based on the
LLmodels \citep{2004A&A...428..993S} grid computed for $T_{\rm eff}=17500$~K,
$\log g=4.0$ and the helium abundance from $\log(\mathrm{He/H})=-3.0$ to
$\log(\mathrm{He/H})=1.9$.

\begin{figure*}

\includegraphics[angle=0,width=12cm,clip=]{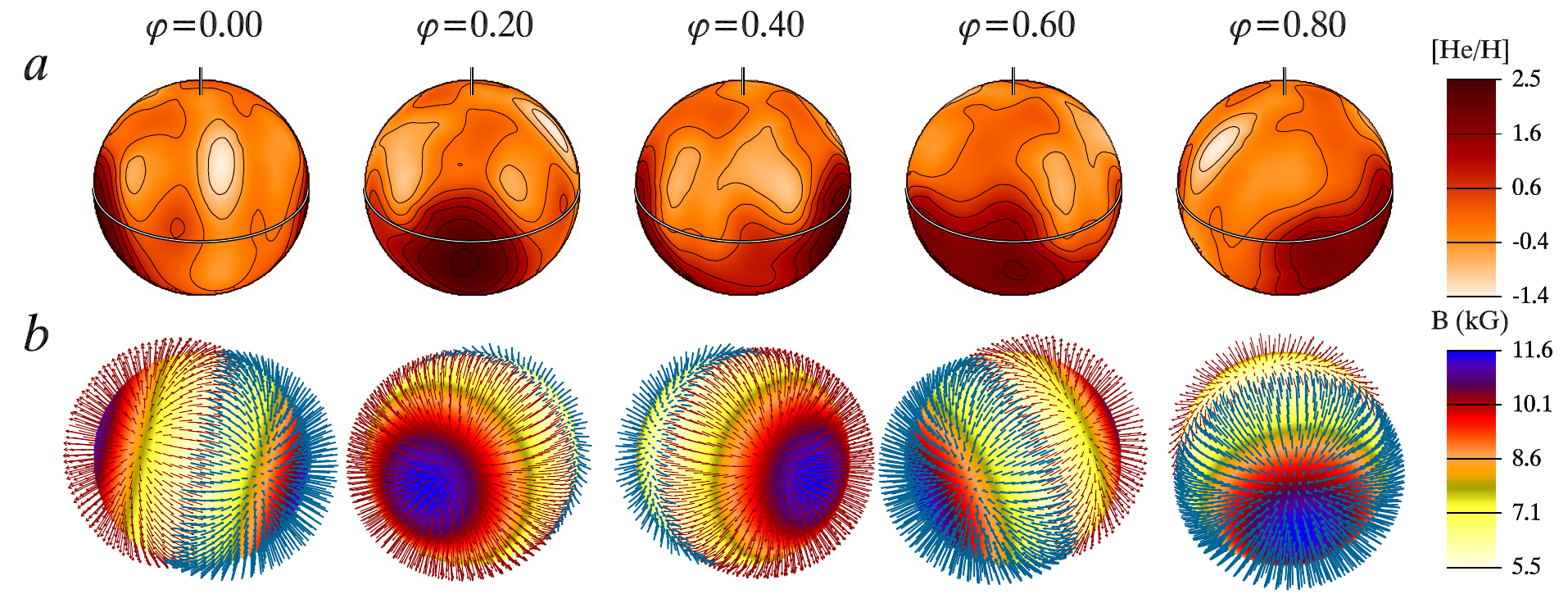}%

\includegraphics[angle=0,width=12cm,clip=]{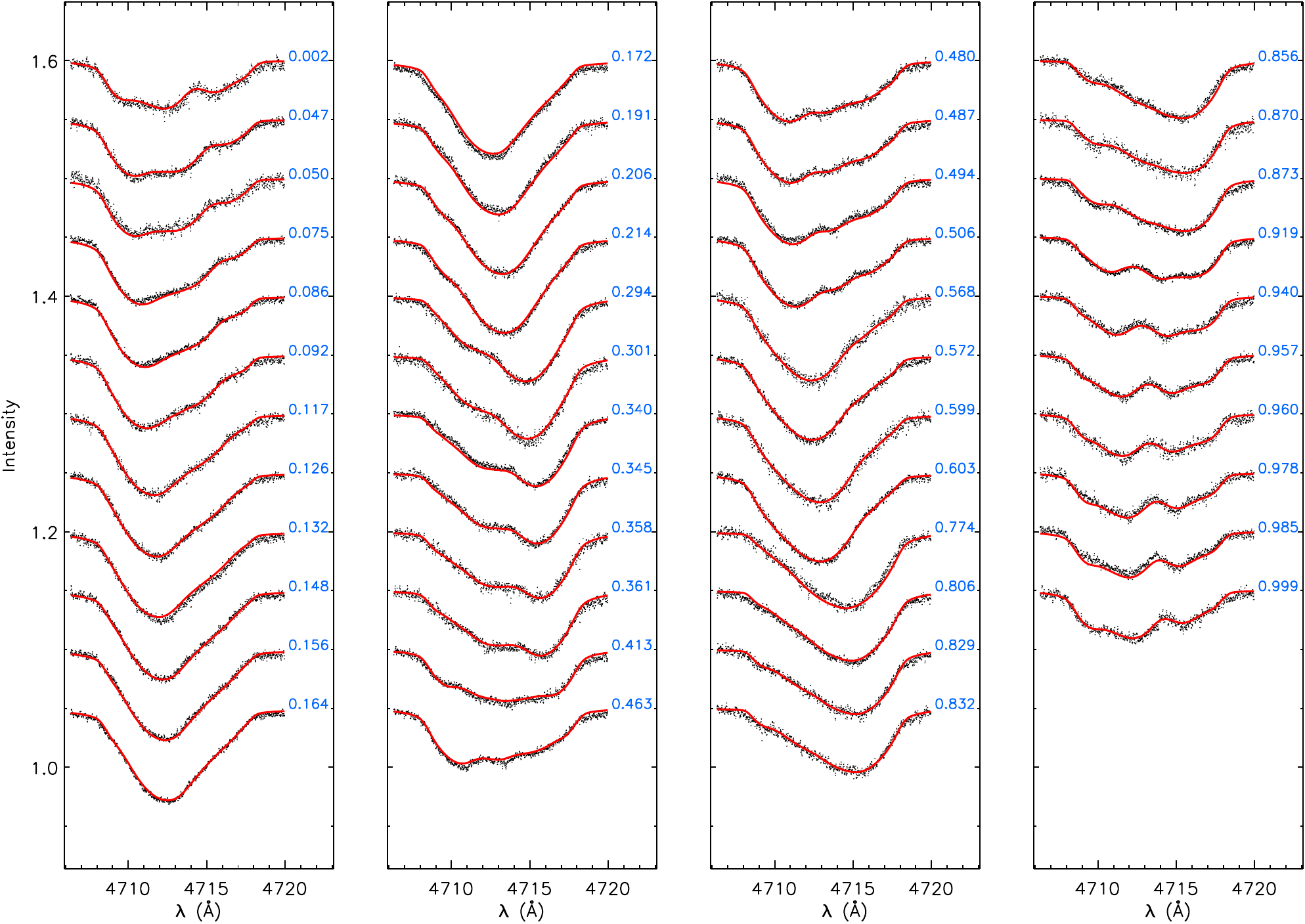}%

\caption[]{Same as Fig.~\ref{fig_DI_CII} for helium distribution reconstructed
  using the He{\sc i}$\lambda$\,4713 line. The He abundance is given in logarithmic
  units relative to the Sun.}
\label{fig_DI_HeI}
\end{figure*}

The results of He abundance inversion are presented in
Fig.~\ref{fig_DI_HeI}. The overall contrast of the He map is nearly 4~dex. The
main He overabundance features are located at the equator and below. To
reproduce a narrow bump traveling across the He line profile at phases
0.95-0.05, the code requires a latitudinally extended spot where He is
strongly underabundant relative to the Sun.

%We used the available spectra for Doppler imaging. Since the magnetic data is
%sufficient only to derive the mean longitudinal field, but actually a magnetic
%Doppler imaging code was used, a dipole with the geometry given in the next
%section was assumed \citep[see][for a description and a preliminary
%  application to HR\,7355]{2011IAUS..272..166K}.
%
%{\bf I guess some more detail should be written here. Oleg, can you please
%  provide something on technical details, while I will write some more on
%  comparison to other stars (like HD37776, sigOri)?}

%%%%%%%%%%%%%%%%%%%%%%%%%%%%%%%%%%%%%%%%%%%%%%%%%%%%%%%%%%%%%%%%%%%%%%%
%%%%%%%%%%%%%%%%%%%%%%%%%%%%%%%%%%%%%%%%%%%%%%%%%%%%%%%%%%%%%%%%%%%%%%%
%%%%%%%%%%%%%%%%%%%%%%%%%%%%%%%%%%%%%%%%%%%%%%%%%%%%%%%%%%%%%%%%%%%%%%%
%%%%%%%%%%%%%%%%%%%%%%%%%%%%%%%%%%%%%%%%%%%%%%%%%%%%%%%%%%%%%%%%%%%%%%%
%%%%%%%%%%%%%%%%%%%%%%%%%%%%%%%%%%%%%%%%%%%%%%%%%%%%%%%%%%%%%%%%%%%%%%%
%%%%%%%%%%%%%%%%%%%%%%%%%%%%%%%%%%%%%%%%%%%%%%%%%%%%%%%%%%%%%%%%%%%%%%%

%%%%%%%%%%%%%%%%%%%%%%%%%%%%%%%%%%%%%%%%%%%%%%%%%%%%%%%%%%%%%%%%%%%%%%%
%%%%%%%%%%%%%%%%%%%%%%%%%%%%%%%%%%%%%%%%%%%%%%%%%%%%%%%%%%%%%%%%%%%%%%%
%%%%%%%%%%%%%%%%%%%%%%%%%%%%%%%%%%%%%%%%%%%%%%%%%%%%%%%%%%%%%%%%%%%%%%%
%%%%%%%%%%%%%%%%%%%%%%%%%%%%%%%%%%%%%%%%%%%%%%%%%%%%%%%%%%%%%%%%%%%%%%%
%%%%%%%%%%%%%%%%%%%%%%%%%%%%%%%%%%%%%%%%%%%%%%%%%%%%%%%%%%%%%%%%%%%%%%%
%%%%%%%%%%%%%%%%%%%%%%%%%%%%%%%%%%%%%%%%%%%%%%%%%%%%%%%%%%%%%%%%%%%%%%%

%%%%%%%%%%%%%%%%%%%%%%%%%%%%%%%%%%%%%%%%%%%%%%%%%%%%%%%%%%%%%%%%%%%%%%%
%%%%%%%%%%%%%%%%%%%%%%%%%%%%%%%%%%%%%%%%%%%%%%%%%%%%%%%%%%%%%%%%%%%%%%%

%%%%%%%%%%%%%%%%%%%%%%%%%%%%%%%%%%%%%%%%%%%%%%%%%%%%%%%%%%%%%%%%%%%%%%%
%%%%%%%%%%%%%%%%%%%%%%%%%%%%%%%%%%%%%%%%%%%%%%%%%%%%%%%%%%%%%%%%%%%%%%%
\section{The magnetic field}
\label{sec_magfield}

%\subsection{Inclination and magnetic obliqueness}

While the high $v \sin i$ tells that the inclination cannot be small, some
further constraints on both the inclination $i$ and the obliqueness of the
magnetic field vs.\ the rotational axis $\beta$ can be obtained from the
observed properties of the longitudinal magnetic field.

The magnetic field curve is not centered around zero, i.e.\ $B_{\rm min} \neq
-B_{\rm max}$, as noted in the original papers \citep{oksala,rivi7355}. If
a dipole is assumed, this means neither $i$ nor $\beta$ can exactly be
90\degr, although they might be close: If the inclination was 90\degr, the
rotational projection of the two magnetic poles must be the same, regardless
of $\beta$. The same is true for $\beta=90\degr$, since then the magnetic
poles would be on the rotational equator.

Although the field, for reasons discussed in Sect.~\ref{sec_discussion}, is
most likely not exactly dipolar, the variation is sufficiently sinusoidal to
apply the equations by \citet{1971PASP...83..571P} to the magnetic variation
given by \citet{rivi7355}. This yields $r = B_{\rm min} / B_{\rm max} =
\cos(\beta+i)/\cos(\beta-i)= -0.78 $. If the values of \citet{oksala} are
used, $r=-0.80$ is obtained, but this changes the values of $\beta$ and $i$
(see below) only by a few degrees. This value of $r$ means that $\beta+i$ must
be smaller than or equal to 141\degr. If the sum is equal to 141\degr, then
$i=\beta=70.5\degr$. If $ i\neq \beta$, we can say that on the one hand, $i$
cannot be much smaller than 70\degr\ because of the large $v\sin i$, on the
other hand $i$ cannot be much larger either, since then $\beta$ would have to
be very small, which would make it very hard to explain the strong mean
longitudinal field, unless the dipole field strength reaches several dozen
kilogauss or more at the poles.  We conclude that both $i$ and $\beta$ are
most likely between 60 and 80\degr, in such a way that $\beta+i$ is about 130
to 140\degr. This limit to the inclination is in nice agreement with the
spectroscopic determination, which is $60\pm10\degr$.

Using \citeauthor{1950ApJ...112..222S}'s (\citeyear{1950ApJ...112..222S}) in
the form given by \citet{2000A&A...355..315L}, and assuming a pure dipole with
$i=60\degr$ and $\beta=75\degr$, the field strength at the magnetic poles
would then be about 11.6\,kG.

The observational indications for a non-dipole field and a non-sinusoidal
field curve are subtle, so that the above values, summarized in
Table~\ref{tab_magpar}, should at least be a good approximation.

%%%%%%%%%%%%%%%%%%%%%%%%%%%%%%%%%%%%%%%%%%%%%%%%%%%%%%%%%%%%%%%%%%%%%%%
%%%%%%%%%%%%%%%%%%%%%%%%%%%%%%%%%%%%%%%%%%%%%%%%%%%%%%%%%%%%%%%%%%%%%%%
%%%%%%%%%%%%%%%%%%%%%%%%%%%%%%%%%%%%%%%%%%%%%%%%%%%%%%%%%%%%%%%%%%%%%%%
\begin{figure*}
\parbox{0.5cm}{~}%
\parbox{6cm}{\centerline{H$\alpha$}}%
\parbox{6cm}{\centerline{H$\beta$}}%
\parbox{6cm}{\centerline{Pa$_{14}$}}%

\includegraphics[viewport=96 156 600 828,angle=0,width=6cm,clip=]{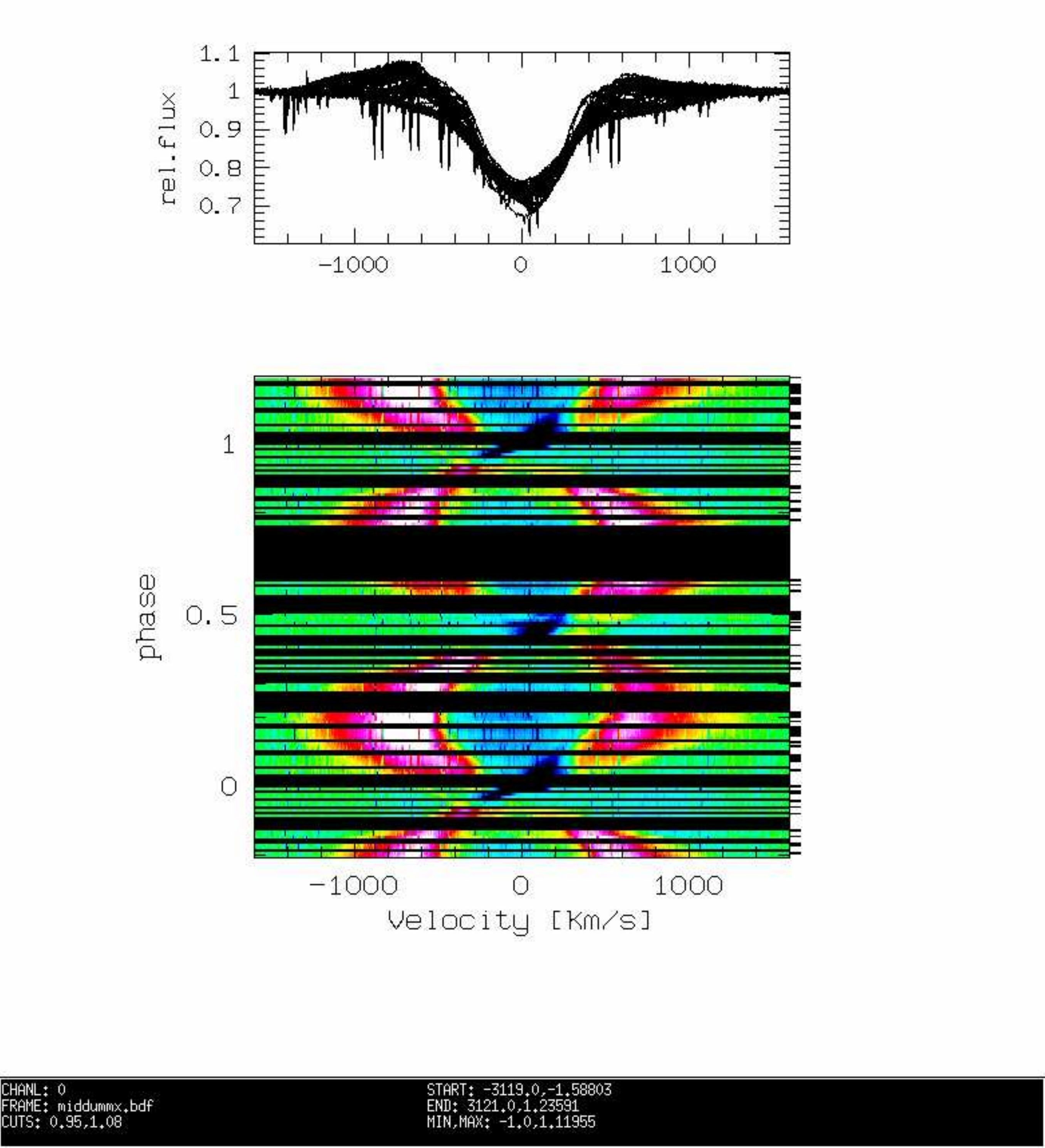}%
\includegraphics[viewport=96 156 600 828,angle=0,width=6cm,clip=]{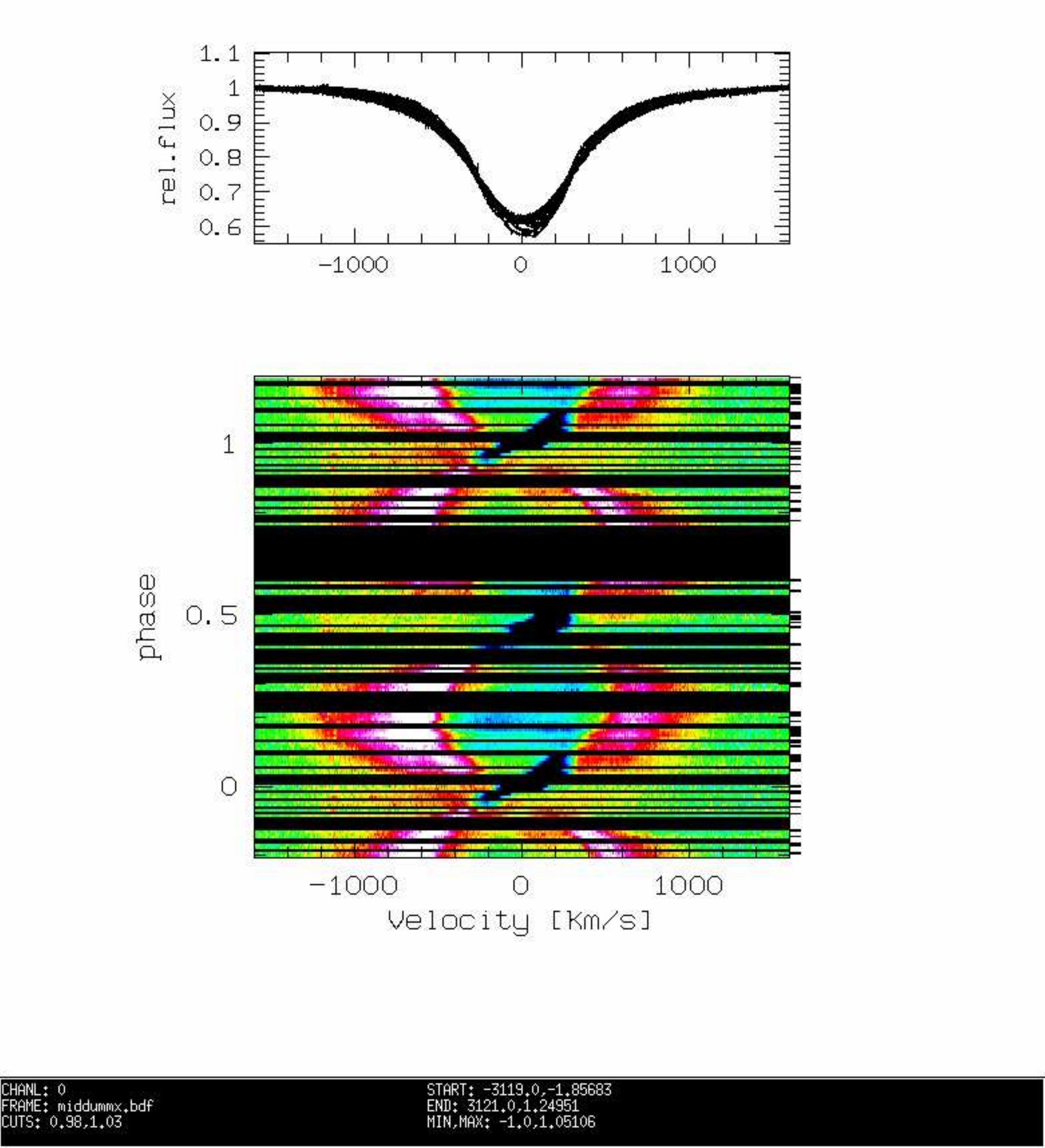}%
\includegraphics[viewport=96 156 600 828,angle=0,width=6cm,clip=]{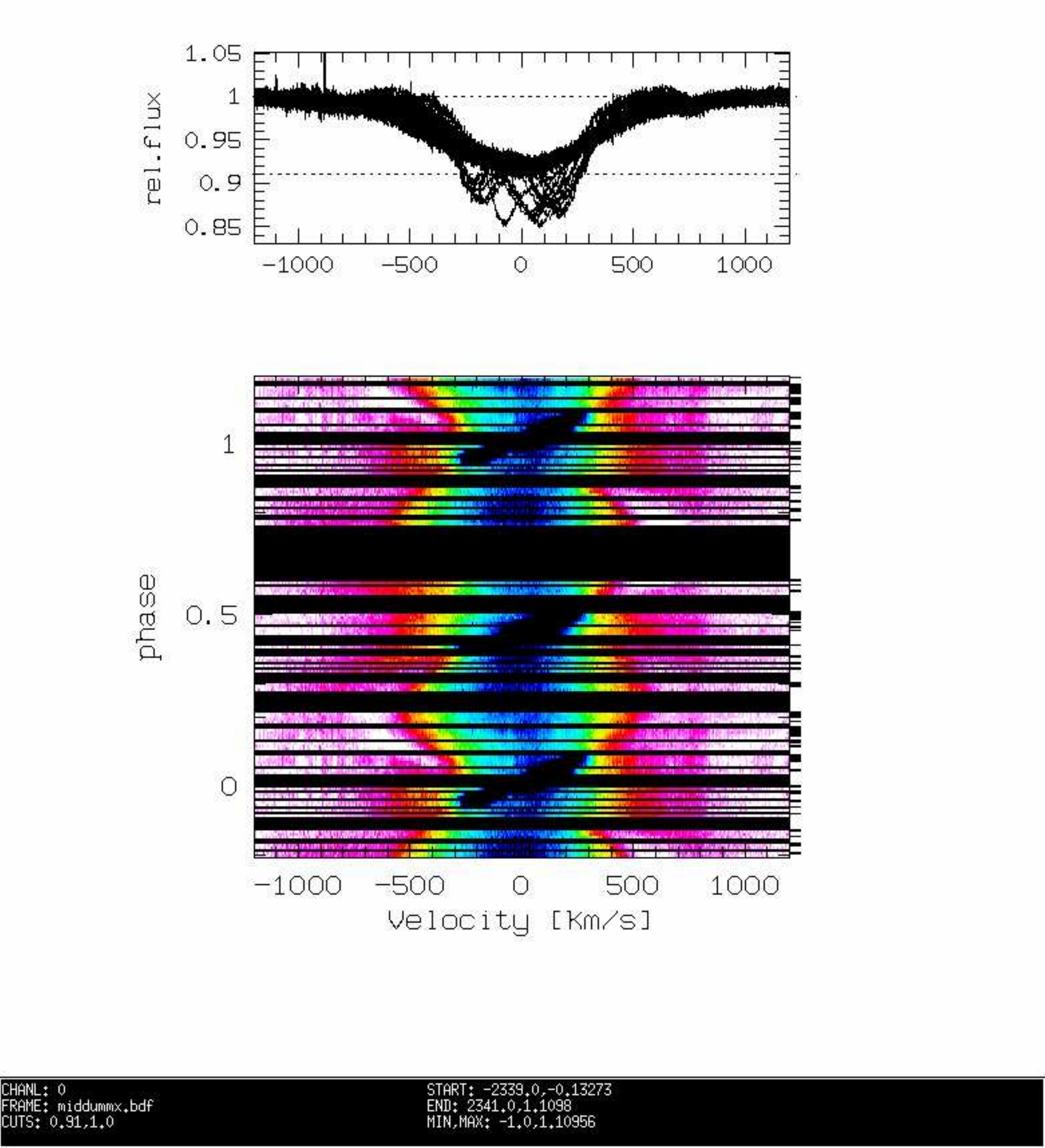}%

\caption[]{Observational variations of the circumstellar magnetosphere. The
  upper row shows the actual profiles overplotted, in the lower row the phased
  variations are shown. For H$\alpha$ and H$\beta$ the residuals after
  subtracting the Bruce3 model are used for the phased plot; the residual
  level in the line center is zero to better than 0.01. Emission is clearly
  seen in the residuals, in fact up to H$\delta$. Paschen lines show emission
  as well; shown here is Pa$_{14}$ (note the different scales of the
  abscissae).  128 phase bins were used for all panels in order to sample the
  fast variations around the occultation phases in sufficient detail.}
\label{fig_dynamicalmagneto}
\end{figure*}
%%%%%%%%%%%%%%%%%%%%%%%%%%%%%%%%%%%%%%%%%%%%%%%%%%%%%%%%%%%%%%%%%%%%%%%
%%%%%%%%%%%%%%%%%%%%%%%%%%%%%%%%%%%%%%%%%%%%%%%%%%%%%%%%%%%%%%%%%%%%%%%
%%%%%%%%%%%%%%%%%%%%%%%%%%%%%%%%%%%%%%%%%%%%%%%%%%%%%%%%%%%%%%%%%%%%%%%

%%%%%%%%%%%%%%%%%%%%%%%%%%%%%%%%%%%%%%%%%%%%%%%%%%%%%%%%%%%%%%%%%%%%%%%
%%%%%%%%%%%%%%%%%%%%%%%%%%%%%%%%%%%%%%%%%%%%%%%%%%%%%%%%%%%%%%%%%%%%%%%
\section{The circumstellar magnetosphere}
\label{sec_magnetosphere}

The circumstellar magnetosphere is filled with magnetically trapped gas
\citep[see e.g. Figs.~1 and 8 of][for an illustration of the concept and
  definitions]{2007MNRAS.382..139T}. The trapped gas, according to
\citeauthor{2007MNRAS.382..139T}, occupies a warped plane between the magnetic
and rotational equator, effectively the points along each field line where the
centrifugal acceleration is highest, as long as this is larger than the local
gravity.  However, this plane is not uniformly filled; the intersections of
the magnetic and rotational equatorial planes are preferred, forming two
geometrically relatively thin lobes \citep{2008MNRAS.389..559T}.  This gas is
seen in emission in hydrogen lines when in quadrature and in absorption in
hydrogen and a few other lines when passing through the line of sight. Species
and lines in which the occultations are present include H{\sc i}, O{\sc
  i}$\lambda$\,8446 and the 777x triplet, Si{\sc ii}$\lambda$\,6347, N{\sc
  ii}$\lambda$ 4631 (see respective panel in Fig.~\ref{fig_dynamicalphoto}, at
$\phi=0.46$ and $v=0$\,km\,s$^{-1}$), and many Fe{\sc ii} lines (Fe{\sc
  ii}$\lambda$\,4949 is visible in the Si{\sc iii}$\lambda$\,4553 panel in
Fig.~\ref{fig_dynamicalphoto}, at $\phi=0.46$ and $v=-200$\,km\,s$^{-1}$).

%%%%%%%%%%%%%%%%%%%%%%%%%%%%%%%%%%%%%%%%%%%%%%%%%%%%%%%%%%%%%%%%%%%%%%%
%%%%%%%%%%%%%%%%%%%%%%%%%%%%%%%%%%%%%%%%%%%%%%%%%%%%%%%%%%%%%%%%%%%%%%%
%%%%%%%%%%%%%%%%%%%%%%%%%%%%%%%%%%%%%%%%%%%%%%%%%%%%%%%%%%%%%%%%%%%%%%%
\begin{table}
\caption[]{\label{tab_magpar}Magnetic parameters. The inclination
  $i=60\pm10\degr$ is taken from Table~\ref{tab_modpar}}.
\begin{center}
\begin{tabular}{lr}
\multicolumn{2}{c}{Field parameters}\\
$\beta$ & $75\pm10\degr$\\
Polar $B_{\rm dipole}$  & 11.6\,kG \\
\multicolumn{2}{c}{Magnetospheric lobe parameters}\\
Inner radius  & 2\,R$_\star$\\
Outer radius  & 4\,R$_\star$\\
Particle density & $10^{12.3\pm0.5} {\rm cm}^{-3}$\\
Plane tilt   & $<22\degr/\cos i$\\
Lobe separation  & $165\pm5\degr$ \\
\end{tabular}
\end{center}
\end{table}
%%%%%%%%%%%%%%%%%%%%%%%%%%%%%%%%%%%%%%%%%%%%%%%%%%%%%%%%%%%%%%%%%%%%%%%
%%%%%%%%%%%%%%%%%%%%%%%%%%%%%%%%%%%%%%%%%%%%%%%%%%%%%%%%%%%%%%%%%%%%%%%
%%%%%%%%%%%%%%%%%%%%%%%%%%%%%%%%%%%%%%%%%%%%%%%%%%%%%%%%%%%%%%%%%%%%%%%

%%%%%%%%%%%%%%%%%%%%%%%%%%%%%%%%%%%%%%%%%%%%%%%%%%%%%%%%%%%%%%%%%%%%%%%
\subsection{Geometry}

The two occultations occur at phases $\phi=0.0$ and $\phi =0.46$
(Fig.~\ref{fig_dynamicalmagneto}). The phenomena associated with the passing
of the magnetospheric lobes in front of the star provide information on the
circumstellar geometry.  The lobe absorption is a relatively narrow feature,
of a FWHM of only about 250\,km\,s$^{-1}$ in Pa$_{14}$. This value is similar
in other lines. This corresponds to about 40\,\% of $2\times(v \sin i)$. Since
the circumstellar gas is locked to the magnetosphere, it is in co-rotation
with the stellar surface. For this reason any projected radial velocity
interval is unambiguously mapped onto a strip parallel to the projected
rotational axis, and the width of such an interval translates directly to
geometry: A $90\degr$ tilt would produce zero width (as only one velocity
would be obscured at any given time), while a $0\degr$ tilt would obscure all
velocities across $2\times(v\sin i)$ simultaneously. Hence, the projection of
the tilt of the plane of the lobes against the rotational axis must then be
around the arcus tangens of the simultaneous velocity dispersion of the
circumstellar absorption in units of $2\times(v \sin i)$, which in this case
is $\arctan 0.4 = 22\degr$. This actually is an upper limit, if a non-zero
scale height for the lobes was assumed, the inclination had to be even
lower. Note, however, that this tilt is not just a projection of $90\degr -
\beta$, since the plane of the lobes is somewhat warped.  Nevertheless, it is
probably close to this value.

An equatorial feature in bound corotation, traversing the line of sight at
photospheric level, would take half a cycle to do so ($\Delta\phi=0.5$). With
increasing distance, the crossing time, i.e.\ the fraction of the orbit within
a line of sight impact parameter $<1$, decreases quadratically. The passing of
the lobe through the line of sight towards the star is much faster than
$\Delta\phi=0.5$: from $-v \sin i$ to $+v \sin i$ takes the feature about
$\Delta\phi=0.13$. i.e.\ about one fourth of the crossing time as in case
material would be present down to the stellar surface. ThereforeCombining this
and the quadratic decrease with distance, it can be concluded that there is no
absorbing material directly above the star, but sufficient density to absorb
is found only at and above a radius of $2\,R_{\star}$.

Outside the occultation phases, the emitting material is seen next to the
stellar disk, and again since it is in magnetically bound corotation there is
a linear relation between rotational velocity and distance from the stellar
surface, meaning material at twice the radius has twice the rotational
velocity. Thus, when the circumstellar material is in quadrature, the observed
velocities give additional geometry information: The emission has an inner
edge at about $600$\,km\,s$^{-1}$, which again points to an empty region
inside 2\,R$_{\star}$. The outer edge of the emission is at a velocity of
about $4\times v \sin i$ in H$\alpha$ and $3\times v \sin i$ in Pa$_{14}$,
which gives outer limits for the lobe emission of about 4\,R$_{\star}$ and
3\,R$_{\star}$, respectively.

There is also a ``skew'' in the appearance of the emitting lobes, well seen in
Fig.~\ref{fig_dynamicalmagneto}. Tracing the maximum emission levels belonging
to each lobe, it is found that each lobe varies in a sinusoidal shape,
i.e.\ the cardinal points of conjunctions and quadratures are separated by a
quarter of a cycle, but the two lobes are separated by less than half a cycle,
i.e.\ geometrically by less than 180\degr. That difference is the same as the
one that causes the second occultation to be at $\phi=0.46$ instead of at
$\phi=0.5$.

%%%%%%%%%%%%%%%%%%%%%%%%%%%%%%%%%%%%%%%%%%%%%%%%%%%%%%%%%%%%%%%%%%%%%%%
%%%%%%%%%%%%%%%%%%%%%%%%%%%%%%%%%%%%%%%%%%%%%%%%%%%%%%%%%%%%%%%%%%%%%%%
%%%%%%%%%%%%%%%%%%%%%%%%%%%%%%%%%%%%%%%%%%%%%%%%%%%%%%%%%%%%%%%%%%%%%%%
\begin{table}
\caption[]{\label{tab_decrements}Equivalent width of the circumstellar
  emission in the Balmer lines and the computed Balmer decrements for two
  phases and the red and blue emission peaks separately.}
\begin{center}
\begin{tabular}{rlllll}
  & \multicolumn{3}{c}{Equivalent widths [\AA ]} & 
    \multicolumn{2}{c}{Balmer decr.} \\
  & H$\alpha$&H$\beta$&H$\gamma$& $D_{34}$&$D_{54}$\\
\hline
Blue, $\phi=0.17$ & $-1.37$ & $-0.42$ & $-0.17$ & 1.16 & 0.59 \\
Red, $\phi=0.17$ & $-0.94$ & $-0.25$ & $-0.13$ & 1.33 & 0.76 \\
Blue, $\phi=0.78$ & $-1.00$ & $-0.33$ & $-0.14$ & 1.07 & 0.62 \\
Red, $\phi=0.78$ & $-1.17$ & $-0.26$ & $-0.08$ & 1.59 & 0.45 \\
\end{tabular}
\end{center}
\end{table}
%%%%%%%%%%%%%%%%%%%%%%%%%%%%%%%%%%%%%%%%%%%%%%%%%%%%%%%%%%%%%%%%%%%%%%%
%%%%%%%%%%%%%%%%%%%%%%%%%%%%%%%%%%%%%%%%%%%%%%%%%%%%%%%%%%%%%%%%%%%%%%%
%%%%%%%%%%%%%%%%%%%%%%%%%%%%%%%%%%%%%%%%%%%%%%%%%%%%%%%%%%%%%%%%%%%%%%%

%%%%%%%%%%%%%%%%%%%%%%%%%%%%%%%%%%%%%%%%%%%%%%%%%%%%%%%%%%%%%%%%%%%%%%%
\subsection{Density}

The theoretical line profiles are a very good approximation of the mean
photospheric profiles, as seen in Fig.~\ref{fig_fluxfit} for H$\gamma$ and
Fig.~\ref{fig_dynamicalmagneto} for H$\beta$ and H$\alpha$. Therefore, the
residuals faithfully represent the pure circumstellar emission and it is
possible to apply nebular diagnostics and calculate Balmer decrements. The
Balmer decrements $D_{34}$ and $D_{54}$ were measured separately for the blue-
and redshifted emission peaks at phases 0.17 and 0.78, when the emitting
material is seen next to the star (see Table~\ref{tab_decrements}). For
computing the decrements, the same method as used by
\citet{2003A&A...402..253S} for the classical Be star 28\,CMa was adopted. The
stellar continuum fluxes to correct for the normalization were those modeled
by Bruce3, so that $f_\star({\rm H}\alpha)/f_\star({\rm H}\beta)=0.35 $ and
$f_\star({\rm H}\gamma)/f_\star({\rm H}\beta)=1.46 $. The measured values were
compared to the theoretical computations by \citet{1988ApJ...326..738W}, which
were derived for an isothermal pure hydrogen accretion disk of 10\,000\,K,
that is optically thin in the continuum. The Balmer decrements depend only
weakly on temperature in such a hot plasma, so that even if the circumstellar
material of HR\,7355 is hotter than 10\,000\,K the derived densities still
provide useful guidelines for modeling.  While the values for $D_{54}$ would
be in agreement with logarithmic particle densities between 11.7 and 12.5 per
cm$^3$, the logarithmic particle densities inferred from $D_{34}$ are somewhat
higher, at 12.2 to 12.8 per cm$^3$. In any case, these values are close to the
optically thick limit, above which the decrements become independent of
density.

%%%%%%%%%%%%%%%%%%%%%%%%%%%%%%%%%%%%%%%%%%%%%%%%%%%%%%%%%%%%%%%%%%%%%%%
%%%%%%%%%%%%%%%%%%%%%%%%%%%%%%%%%%%%%%%%%%%%%%%%%%%%%%%%%%%%%%%%%%%%%%%
\section{Discussion}
\label{sec_discussion}
\subsection{HR\,7355 as a rapid rotator}

HR\,7355 was the first magnetic star discovered to rotate so rapidly that
rotational effects like oblateness and gravity darkening should be taken into
account. A second example, HR\,5907, has been published by
\citep{2012MNRAS.419.1610G}. Due to the relative proximity of HR\,7355 not
only was it possible to obtain high quality data, but the public archives
contain data for the star as well.

Using all these data, the five parameters required to describe a rapidly
rotating star could be derived (Table~\ref{tab_modpar}). Of these five, two
stand out: Firstly, the effective temperature of just 17\,500\,K. This is
fairly low for a spectral classification of B2. We note, however, that B2 is
the spectral subtype at which He{\sc i} lines are strongest, and so the
classification for a star with helium overabundance will be biased towards
that spectral subtype with respect to its temperature. Although the Doppler
imaging indicates that not the entire surface is overabundant in helium; with
patches of both over- and underabundance existing, the stellar surface on
average is overabundant in helium, justifying its classification as He-strong
in earlier papers \citep{oksala,rivi7355}.

The second unusual parameter is the rotation rate, at almost 90\% of the
critical value (even the previous record holder for rapid rotation among
non-degenerate magnetic OB stars, $\sigma$\,Ori\,E barely reached 50\%).  This
means that no significant rotational spin-down can have occurred for HR\,7355
yet. Unless the field is self-generated and has only existed since recent
times, this leaves two possibilities: Either the star is so young that it had
no time to brake yet, or the mass loss is so weak that it does not remove any
significant amount of angular momentum. While the latter would have been
unlikely for a star of 22\,000\,K, i.e.\ corresponding to a normal B2 star, it
is much more plausible for a star of only 17\,500\,K. For the very similar
HR\,5907, \citet{2012MNRAS.419.1610G} obtain a spin-down timescale of
8\,Myr. This is still less than the 15 to 25\,Myr suggested by
\citet{2010A&A...511L...7M}. However, for an age of less than 8\,Myr our
stellar parameters, i.e.\ gravity, mass, polar radius, and luminosity, agree
with evolutionary models. Even at a distance of only 236\,pc the conclusion
by \citet{2010A&A...511L...7M}, namely that HR\,7355 is probably not a member
of Scorpius-Centaurus, stands.

\subsection{Magnetic field geometry}

While for a photospheric parameter analysis the tools exist to take
rotation into account, such tools for the magnetic analysis are only being
developed and have not yet reached the parameter space required here, simply
because the inclusion of such strong effects was not necessary to date. The
most rapid rotation taken into account for magnetic analysis so far was by
\citet{2005A&A...429..677K}, limited to $R_{\mathrm{pole}}/R_{\mathrm{eq}} <
0.95$.  In the following discussion, the effects of the rapid
rotation are speculated about, but we stress firm conclusions will require
more sophisticated theoretical work to be done.

A sinusoidal variation of the longitudinal magnetic field is the hallmark of a
centered dipole field tilted with respect to the rotation axis. However, the
high-quality spectroscopic data suggest that the field topology may be more
complex than this. With the circumstellar clouds situated at the intersections
between magnetic and rotational equators, the cloud transits should be
coincident with $B_z=0$.  The measured magnetic curve, when fitted with a
sine, has an offset of $+0.3$\,kG. The magnetic nulls are expected to occur at
$\phi=0$ and $\phi=0.54$ (dotted lines in lowermost panel of
Fig.~\ref{fig_phasingall}). However, the cloud transits are observed to occur at
$\phi=0.0$ and $\phi=0.46$ (dotted lines in upper panels of
Fig.~\ref{fig_phasingall}).  The maximal strength of the He{\sc i} lines
normally coincides with the magnetic poles pointing towards the observer. This
would be expected for $\phi=0.28$ and $0.78$ in case of a sinusoidal variation
(dashed lines in lowermost panel of Fig.~\ref{fig_phasingall}), but
spectroscopically, i.e.\ estimated by maximal He-enhancement, occurs at
$\phi=0.25$ and $0.68$ (dashed lines in upper panels of
Fig.~\ref{fig_phasingall}).

Offsets as discussed in the previous poaragraph could theoretically be
explained in case of a de-centered magnetic dipole. The two misaligned cloud
crossings would require an offset of the dipole along the magnetic polar axis.
Such an offset could also account for the constant term of 0.3\,kG in the
field curve, although this alone could also be explained by the projection
topology of an aligned field.  The observed timing of the magnetic poles
facing towards the observer requires an offset perpendicular to this
axis. These offsets would show in a non-sinusoidal variability curve of the
magnetic field. In any case, however, off-center dipoles, or any combination
of low order multipole components, can only be a simplified approximation to
the actual magnetic field, as magnetic Doppler imaging has shown that more
complex and small scale structures are usually present even for stars with
largerly sinusoidal variations in the longitudinal
field \citep{2011IAUS..272..166K}.

%\subsection{Line profile variations} 
The line profile variations are surprisingly complex, with at least four,
possibly as many as six subfeatures identifiable in some lines.  It is
surprising that a relatively strong and ordered field as in HR\,7355 gives
rise to so many clearly distinct zones of helium and metal abundances. One
possibility to be explored could be an interplay of a more simple type of
abundance variation pattern with the co-latitude dependent line formation in a
rapidly rotating star due to the gravity darkening.

\subsection{The magnetosphere}
The circumstellar magnetosphere is filled with dense gas, sufficiently dense
to make the Balmer emission optically thick. Since the gas needs to be ionized
to be trapped by the magnetic field, one might wonder whether in the central
regions of this dense gas this is still the case. The occultation pattern
observed in other line transitions can be used for guidance to answer this
question: The signature of the occultation of the central, most dense part of
the magnetosphere is seen in lines of {Fe}{\sc ii}, N{\sc ii}, Si{\sc ii}
etc. It is not seen, however, in the lower excitation/ionization lines of
Ca{\sc ii} or Na{\sc i}. This means that the temperature of the magnetosphere
is still of the order of 10\,000\,K, where hydrogen is still far more than
sufficiently ionized to be locked to the magnetic field.

\subsection{Comparison to  $\mathbf{\sigma}$\,Ori E and HR\,5907}

The prototype of an emission line magnetosphere star is ${\sigma}$\,Ori E. It
is, in fact, fairly similar to HR\,7355: It is He-strong, with similar polar
field strength and $i=75\degr$ and $\beta=55\degr$. The two main differences
concerning parameters are the higher effective temperature (23\,000\,K) and
slower rotation (less than 50\% critical) of ${\sigma}$\,Ori E \citep[see
  e.g.][for the most recent parameter determination]{2012MNRAS.419..959O}. As
a consequence of geometry and field strength, the circumstellar variations are
comparable, except for the lower $v \sin i$. Fig.~5 of
\citeauthor{2012MNRAS.419..959O} is extremely similar to our
Fig.~\ref{fig_dynamicalmagneto}. As well, occultation crossings can be seen in
metal ions of ${\sigma}$\,Ori E, as in Fig.~6 of that work, Fe{\sc
  iii}$\lambda$\,5127 at $\phi=0$, vs.\ our Fig.~\ref{fig_dynamicalphoto},
N{\sc ii}$\lambda$\,4631 at $\phi=0.46$. A morphological difference, for which
it is not clear how it is associated with the parameters, is that the
equivalent width changes due to the surface chemical abundance pattern are of
much lower amplitude in ${\sigma}$\,Ori E than in HR\,7355. 
%Since the
%geometry is fairly similar, the amplitude of the abundance variations is
%probably smaller than for HR\,7355, but the underlying physical processes are
%not well known.

Shortly after HR\,7355 was confirmed to be a magnetic star
\citep{oksala,rivi7355}, a very similar object with an even shorter period was
found, HR\,5907 \citep{2011IAUS..272..190G,2012MNRAS.419.1610G}. Mass,
effective temperature, and inclination of HR\,7355 and HR\,5907 are within
each other's error bars (see Table~\ref{tab_modpar}), and the period as well
hardly differs with 0.521\,d vs.\ 0.508\,d. The equatorial velocity of
HR\,5907 is somewhat slower, and the equatorial radius smaller than for
HR\,7355, so that HR\,5907 is less luminous, but the average photospheric
spectra of HR\,7355 and HR\,5907 are virtually identical. Also the dipolar
field strength for both is of the order of 10\,kG, and the gas density of the
magnetosphere is comparable as well.

The only strong difference is the obliquity of the field, being much smaller
for HR\,5907 (almost aligned, $\beta=7\degr$), which, however, has striking
effects on the variability. The amplitude of the line profile variability is
much smaller in HR\,5907, since the magnetic equatorial belt is visible to the
observer all the time, and the contributions from the pole only vary slightly,
while for HR\,7355 magnetic pole-on and equator-on views alternate.
Nevertheless the actual profile variability is still much more complex than in
slower rotators, even if the equivalent width amplitude is very small. As
pointed out by \citet{2008MNRAS.389..559T}, the low obliquity also leads to
only one occultation per rotational cycle, while HR\,7355 has two. As well due
to this the emission variability of HR\,5907 is not a clear double-helix
pattern as for HR\,7355.

%%%%%%%%%%%%%%%%%%%%%%%%%%%%%%%%%%%%%%%%%%%%%%%%%%%%%%%%%%%%%%%%%%%%%%%
%%%%%%%%%%%%%%%%%%%%%%%%%%%%%%%%%%%%%%%%%%%%%%%%%%%%%%%%%%%%%%%%%%%%%%%
\section{Conclusions}
\label{sec_conclusions}
We have shown that HR\,7355 is an early B type magnetosphere star. In
principle similar to the well known $\sigma$\,Ori\,E, it stands out as the
non-degenerate magnetic star rotating most closely to the critical limit, at
about 90\%, followed by HR\,5907.  The magnetic field of HR\,7355 has a polar
strength of about 11 to 12\,kG, with a morphology close to, but as suggested
by the asymmetries in the equivalent width curves probably not exactly
dipolar, and with a fairly large obliquity of $75\degr$, seen at an inclination
of about $60\degr$.

The stellar parameters point to a relatively young star with an effective
temperature typical for a mid B rather than an early B type star, namely
17\,500\,K. As the typical mass-loss rate for such a temperature is
significantly less than for a B2 type star, the spin-down timescale is several
Myr, and hence the age and rapid rotation are not inconsistent with the
presence of a fossil magnetic field.

The stellar surface shows strong abundance variations, which have been Doppler
imaged for helium and carbon in this work.  Although some small surface areas
in the Doppler imaging have a helium abundance lower than the solar value,
helium enriched areas dominate. Indeed, compared to a model with typical solar
or B-star abundances, the {He}{\sc i} lines are stronger in the time averaged
profile. This is also the explanation for the spectral classification as type
B2 in spite of the effective temperature of only 17\,500\,K.

As recently as ten years ago, $\sigma$\,Ori\,E was the only B-type magnetic
star with an emission line magnetosphere, and $\theta^1$\,Ori\,C was the only
such O star. In the last few years this situation has changed, and HR\,7355 is
not only a new member of the growing class of magnetosphere emission line
stars, but has doubled the parameter space for these objects in the rotational
dimension. Density estimates for the magnetospheric region are similar in all
three B stars mentioned ($\sigma$\,Ori\,E, HR\,5907, HR\,7355).

\section*{acknowledgments}
OK is a Royal Swedish Academy of Sciences Research Fellow, supported by grants
from Knut and Alice Wallenberg Foundation and Swedish Research Council.

The IUE data presented in this paper were obtained from the Multimission
Archive at the Space Telescope Science Institute (MAST). STScI is operated by
the Association of Universities for Research in Astronomy, Inc., under NASA
contract NAS5-26555. Support for MAST for non-HST data is provided by the NASA
Office of Space Science via grant NAG5-7584 and by other grants and contracts.

%%%%%%%%%%%%%%%%%%%%%%%%%%%%%%%%%%%%%%%%%%%%%%%%%%%%%%%%%%%%%%%%%%%%%%%%%
\bibliographystyle{mn2e}
\bibliography{HR7355a,HR7355b}

\end{document}